\documentclass[%
preprint,
 amsmath,amssymb,
 aps,
]{revtex4-2}

\usepackage{graphicx}
\usepackage{dcolumn}
\usepackage{bm}

\usepackage{subcaption}
\usepackage{amssymb}
\usepackage{physics}
\usepackage{siunitx}
\usepackage{hyperref}
\hypersetup{
    colorlinks=true,
    linkcolor=blue,
    filecolor=magenta,      
    urlcolor=cyan,
    pdftitle={Overleaf Example},
    pdfpagemode=FullScreen,
    }

\newcommand{\ba}{\bm{a}}

\newcommand{\be}{\bm{e}}

\newcommand{\br}{\bm{r}}

\newcommand{\bv}{\bm{v}}

\newcommand{\bx}{\bm{x}}

\newcommand{\bR}{\bm{R}}

\newcommand{\bkappa}{\boldsymbol{\kappa}}

\begin{document}

\preprint{APS/123-QED}

\title{Topological Localized Modes In Moir\'{e} Lattices of Bilayer Elastic Plates With Resonators}

\author{Tamanna Akter Jui}
\author{Raj Kumar Pal}%
 \email{rkpal@ksu.edu}
\affiliation{%
Department of Mechanical and Nuclear Engineering, Kansas State University, Manhattan, KS 66506
}%




\date{\today}

\begin{abstract}
We investigate the existence of higher order topological localized modes in moir\'{e} lattices of bilayer elastic plates. Each plate has a hexagonal array of discrete resonators and one of the plates is rotated an angle ($21.78^\circ$) which results in a periodic moir\'{e} lattice with the smallest area. The two plates are then coupled by inter-layer springs at discrete locations where the top and bottom plate resonators coincide. Dispersion analysis using the plane wave expansion method reveals that a bandgap opens on adding the inter-layer springs. The corresponding topological index, namely fractional corner mode, for bands below the bandgap predicts the presence of corner localized modes in a finite structure. Numerical simulations of frequency response show localization at all corners, consistent with the theoretical predictions. The considered continuous elastic bilayered moir\'{e} structures opens opportunities for novel wave phenomena, with potential applications in tunable energy localization and vibration isolation. 
\end{abstract}

\maketitle


\section{Introduction}\label{sec:sec1}
The study of architected two-dimensional ($2D$) moir\'{e} lattice structures has gained a lot of attention, particularly in $2D$ materials. Moir\'{e} lattices are formed when one periodic lattice is rotated with respect to another identical lattice, see Fig.~\ref{fig:rot1} for an example. At specific angles of rotation/twist, a lattice with a larger periodicity results, called the moir\'{e} lattice. Their dispersion surfaces have unique features like flat bands, and nonlinear (interacting) inter-layer coupling effects that enable various exotic phenomena~\cite{dos2007graphene,trambly2010localization,morell2010flat,bistritzer2011moire,wang2012fractal}. Notable examples include recent breakthroughs with twisted bilayer graphene, including high-temperature super-conductivity~\cite{cao2018unconventional} and two-dimensional magnetism~\cite{gonzalez2017electrically}.

These recent discoveries in quantum mechanics have inspired the quest for novel wave phenomena with moir\'{e} structures in diverse physical domains. The ability to independently engineer the rotation angle and inter-layer interactions, combined with advances in fabrication have opened a rich design space. Examples in photonics include flat bands using a hexagonal array on silicon nanodisk ~\cite{dong2021flat}, lasing by semiconductor membrane with a triangular pattern of nanoholes ~\cite{mao2021magic}, topologically protected corner modes~\cite{oudich2021photonic}, and localization-delocalization transition of light ~\cite{wang2020localization}. These phenomena arise solely due to the relative rotation between two lattices, without introducing any structural defect, material discontinuity, or non-linearity. Similarly, in acoustics, bilayer moir\'{e} structure made of coupled acoustic cavities in various lattice configurations have been investigated. It has led to higher-order topological states (HOTI) with hexagonal lattice~\cite{wu2022higher}, acoustic valley edge modes with triangular~\cite{zheng2022topological} and topological Lifshitz transition with square lattice~\cite{yves2022topological}. 

In elastic media, the presence of both longitudinal and shear (transverse) waves offers rich possibilities for novel dynamic phenomena with architected structures. Recent studies have investigated the dynamic properties of moir\'e lattices comprising of elastic plates with arrays of pillars in various configurations. Notable predictions include the existence of non-trivial topological bandgap supporting edge states~\cite{jin2020topological}, chirality-driven flat bands analogous to twisted bilayer graphene~\cite{lopez2020flat,lopez2022theory} and localized modes~\cite{marti2021dipolar}. 
Oudich et al.~\cite{oudich2022twisted} systematically examined the effect of inter-layer coupling in bilayer pillared elastic plates in a twisted honeycomb arrangement. Their calculations predict that a weak coupling gives a dispersion band structure similar to the classical bilayer graphene, while a stronger coupling induces the valley Hall effect for elastic wave propagation. Ruzzene and coworkers studied single-layer elastic plate moir\'{e} structures having a square array of pillars with spatial modulation of heights. They demonstrated  a  topological transition of isofrequency contour and highly directional wave tunability~\cite{yves2022moire}. A majority of these studies have been conducted on large lattices which are difficult to fabricate. In addition, the nonlinear properties of these lattices remain unexplored. The exotic properties of moire structures in electronic media listed above are associated with non-linear or interaction effects of spectrally isolated bands and localized modes. Hence similar localized modes may give rise to analogous nonlinear phenomena in elastic media.

Although localized modes have been extensively investigated in architected structures over the last few decades, recent research has focused on modes that arise due to nontrivial dispersion band topology. 
Their topological origin guarantees their existence and they are immune to structural defects and imperfections. In contrast to accidental or trivial localized modes, topological modes translate across geometric parameters, length scales and material properties. 
Hughes and coworkers~\cite{benalcazar2017quantized,benalcazar2019quantization} developed the theory to establish the topological nature of localized modes at corners and point defects in higher dimensional lattice structures and derived the invariants to systematically infer their presence. Topological localized modes have been observed in diverse physical domains, including  photonic ~\cite{peng2022higher,li2022experimental,xiong2022topological,wang2021higher,wu2021all,proctor2020robustness,oudich2021photonic,mizoguchi2020square}, acoustic~\cite{wang2021straight,deng2022observation,yang2020helical,xue2019acoustic,ni2019observation}, phononic~\cite{serra2018observation,ezawa2018higher,chen2022topology}, and elastic~\cite{chen2021corner,danawe2021existence,ma2023tuning,liu2023second}. They have also been predicted in moir\'e lattices of twisted bilayer graphene with various inter-layer potentials, however there is disagreement between the various predictions on their locations. Liu et al.~\cite {liu2021higher} show the localized mode only at the $120^\circ$ corner and provide symmetry-based reasons for non-existence of such modes at a $60^\circ$ corner, while Wu et al. predicts them at both $60^\circ$ and $120^\circ$ corners~\cite{wu2022higher}.

Here, we investigate the existence of such topological corner localized modes in bilayer elastic moir\'{e} plates. We consider two elastic thin plates having a hexagonal array of resonators. The plates are rotated an angle ($21.78^\circ$) relative to each other to generate a moir\'e pattern. Discrete inter-layer springs are added between the plates at locations where the top and bottom plate resonators coincide. The dispersion spectrum and fractional corner mode are determined for a unit cell to predict the existence of corner localized modes. The predictions are verified through numerical simulations of frequency response on a finite plate. The outline of this paper is as follows: section~\ref{sec:sec2} presents the lattice configuration and the governing equations, followed by a dispersion analysis and computation of topological indices for a unit cell in section~\ref{sec:sec3}. The numerical results of mode shapes and frequency response are presented in section~\ref{sec:numer} and the results are summarized in section~\ref{sec:conc}. 

\section{Lattice description and problem setup}\label{sec:sec2}
We first derive the conditions that result in a periodic hexagonal moir\'e lattice, determine the smallest such lattice and its lattice vectors. Then the elastic plate configuration and its governing equations are presented. 

\subsection{Lattice and unit cell: geometric description}\label{sec:sec2.1}
\begin{figure}[!htbp]
        \centering    
        \includegraphics[height=9cm]{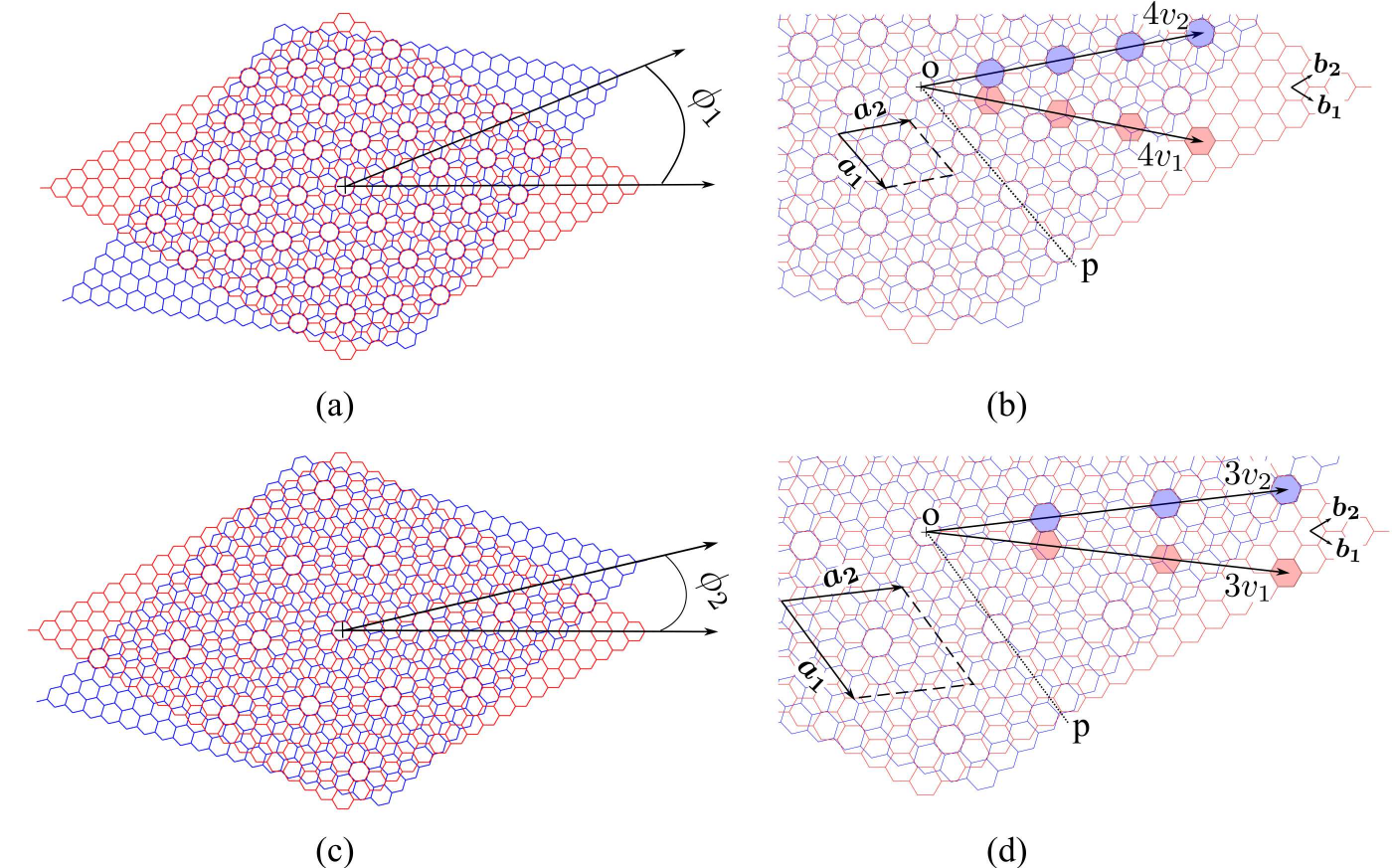}
          \caption{Schematic of periodic moir\'{e} lattices and unit cell. The rotation angle is $\phi_1 = 21.78^{\circ}$ in (a) and $\phi_2 = 13.17^{\circ}$ in (c). (b,d) Enlarged view of the lattices, along with lattice vectors. The blue and red shaded hexagons overlap when the two lattices coincide (at $\phi=0$). The moir\'e lattice in (a,b) has the smallest unit cell. 
          } 
  \label{fig:rot1}
\end{figure}
Figure~\ref{fig:rot1} displays examples of hexagonal moir\'e lattices, along with their unit cells and lattice vectors. They are formed by stacking two identical hexagonal lattices with a relative rotation between them. The rotation is about the center of a hexagon with respect to its out-of-plane axis. The blue and red lattices are identical, but rotated relative to each other about $O$. The lattice vectors of the red hexagonal lattice are $\bm{b_1}$, $\bm{b_2}$ with $60^\circ$ angle between them, and the unit cell length is $b$. 
For an arbitrary relative rotation, the resulting pattern is not periodic. At specific rotation angles, a periodic pattern does result. These angles are hereby called moir\'{e} angles. 

Let us discuss the conditions under which a periodic moir\'e lattice arises. We analyze the configuration that results when the blue lattice, which is initially coincident with the red one, is rotated about $O$. Two videos are presented in the supplementary materials on this rotation, illustrating the formation of the two distinct moir\'{e} lattices of Fig.~\ref{fig:sch}(a,c). Let us fix this rotation center $O$ as the origin of our coordinate system. The key observation is that a periodic moir\'e lattice results when the center of a hexagon in a blue lattice coincides with the center of another hexagon in a red lattice away from the origin. 
The distance between the hexagon center and the rotation center $O$ should be identical for a pair of hexagons, one each from the blue and red lattice. Let us consider a hexagon in the red lattice with center $\bv_1$ at
\begin{equation}
    \bm{v_1} = m \bm{b_1} + n \bm{b_2}, \;\;\; m,n \in \mathbb{Z},\;\;\; m > n,\;\;\; \gcd(m,n) = 1 .  
\end{equation}
Its distance from the center is $\|\bm{v_1} \| = b \sqrt{m^2+mn+n^2}$. It is the nearest red shaded hexagon from the center in the examples in Figs.~\ref{fig:rot1}(b,d). Due to the $C_6$ (6-fold rotation) symmetry of the hexagonal lattice, there are multiple hexagons in the blue lattice at the same distance. A simple choice for a hexagon in the blue lattice is $\bm{v_2} = n \bm{b_1} + m \bm{b_2}$, which satisfies $\|\bm{v_1}\|=\|\bm{v_2}\|$.  
The rotation angle (moir\'e angle) $\phi$ is thus the angle between $\bm{v_1}$ and $\bm{v_2}$, given by 
\begin{equation}
    \cos \phi = \dfrac{\bm{v_1.v_2}}{ \|\bv_1\| \|\bv_2\| } = \dfrac{m^2/2 + n^2/2 + 2mn}{m^2 + n^2 + mn} . 
\end{equation}

Let us see why the resulting bilayered (moir\'e) lattice is periodic along 2 directions and derive the lattice vectors of its unit cell. Note that any integer multiple of $\bm{v_1}$, i.e., $q\bm{v_1}$ is also the center of a hexagon in the red lattice. In addition, this hexagon goes to $q\bm{v_2}$ after rotation, as the angle between $q\bm{v_1}$ and $q\bm{v_2}$ is also $\phi$. Thus the lattice is periodic along $\bm{v_2}$, with periodicity $\|\bm{v_2} \|$. Since both the hexagonal lattices have $C_6$ (6-fold rotation) symmetry about $O$, the combined lattice also has $C_6$ symmetry about $O$. The moir\'e lattice is thus also periodic along directions at angle $\pi/3$ from $\bm{v_2}$. We take its  lattice vectors to be $\bv_2$ and a vector at angle $-\pi/3$ from $\bv_2$. In terms of the hexagonal lattice vectors, the moir\'e lattice vectors $(\ba_1, \ba_2)$ may be expressed as 
\begin{equation}\label{lattice_vecs}
    \ba_1 = n (\bm{b_1} - \bm{b_2}) + m \bm{b_1} = (m+n)\bm{b_1} - n \bm{b_2} , \quad \bm{a_2} = n \bm{b_1} + m \bm{b_2} . 
\end{equation}

Figure~\ref{fig:rot1}(a,c) displays the periodic moir\'e lattices for $(m,n) = (2,1)$ and $(3,2)$. Their corresponding unit cells and lattice vectors are indicated in Fig.~\ref{fig:rot1}(b,d). The relative angles between the blue and red lattices for these lattices are $\phi_1=21.78^{\circ}$ and $\phi_2=13.17^{\circ}$. 
The blue and red shaded hexagons coincide when there is no relative rotation between the two lattices. As the blue lattice is rotated, the blue shaded hexagons move to the locations illustrated in the figure, and they lie along $\bm{v_2}$. The lattice vector $\bm{a_1}$ lies along the line labeled $OP$. 
Note from these examples that the unit cell size of the resulting lattice is, in general, different for different $\phi$ values. 

In this work, we investigate the behavior of the lattice with the smallest moir\'e unit cell, due to its potential ease of fabrication with macro-scale components. To determine this unit cell, let us calculate the unit cell area $A$ for a lattice with unit vectors given by Eqn.~\eqref{lattice_vecs}. It is given by 
\begin{equation}
    A = \|  \bm{a_1} \times \bm{a_2} \| = \dfrac{\sqrt{3} b^2}{2}\left( m^2 + mn + n^2 \right) 
    = \dfrac{\sqrt{3}b^2}{4} \left( m^2 + n^2 + (m+n)^2 \right) . 
\end{equation}
For $m$ and $n$ distinct non-zero integers, a direct calculation shows that $A=7\sqrt{3} b^2/2$ is the minimum area for $m=2$, $n=1$. The lattice vectors are thus 
\begin{equation}
    \bm{a_1} = 3\bm{b_1} - \bm{b_2}, \;\;\; \bm{a_2} = \bm{b_1} + 2 \bm{b_1}. 
\end{equation}

Let us discuss the key properties of this lattice. The parallelogram with lattice vectors labeled in Fig.~\ref{fig:rot1}b displays the chosen unit cell, with the center of the hexagon at its center. This choice ensures that a finite-sized hexagon-shaped lattice will have 6-fold rotation symmetry. This property will be used later in Sec.~\ref{sec:Qcalc} to predict localized modes at corners. This unit cell has 14 nodes of the hexagonal lattice in each layer. Indeed, note that the underlying hexagonal lattice has 2 nodes per unit cell and its area is $\sqrt{3}b^2/2$. Comparing with the moir\'e unit cell area, we see that the latter is 7 times larger and it thus has 14 nodes. In addition, there are two nodes in each unit cell where the red and blue lattices coincide. These nodes are indicated by green circles in the inset of Fig.~\ref{fig:sch}a. Their locations within the unit cell, with respect to the lower $60^\circ$ in this inset, are given by
\begin{equation}
    \bm{p_1} = \bm{b_1} + \dfrac{\bm{b_1}+\bm{b_2}}{3}, \;\;\; \bm{p_2} =2\bm{b_1} + \dfrac{2}{3} \left(\bm{b_1}+\bm{b_2}\right) .
\end{equation}
By checking explicitly, we note that $\bm{p_1}$ and $\bm{p_2}$ lie at different sub-lattice sites of each hexagonal lattice. In particular, $\bm{p_1}$ lies at the $\alpha$ ($\beta$) site in the red (blue) lattice, while $\bm{p_2}$ lies in the $\beta$ ($\alpha$) site. Thus the $\alpha$ ($\beta$) site of the red (blue) lattice coincides with the $\beta$ ($\alpha$) site of the red lattice at $\bm{p_1}$ ($\bm{p_2}$) in each unit cell.

\subsection{Plate configuration and governing equations}\label{sec:plateGovEqn}
\begin{figure}[!htbp]
        \centering    
        \includegraphics[height=4.5cm]{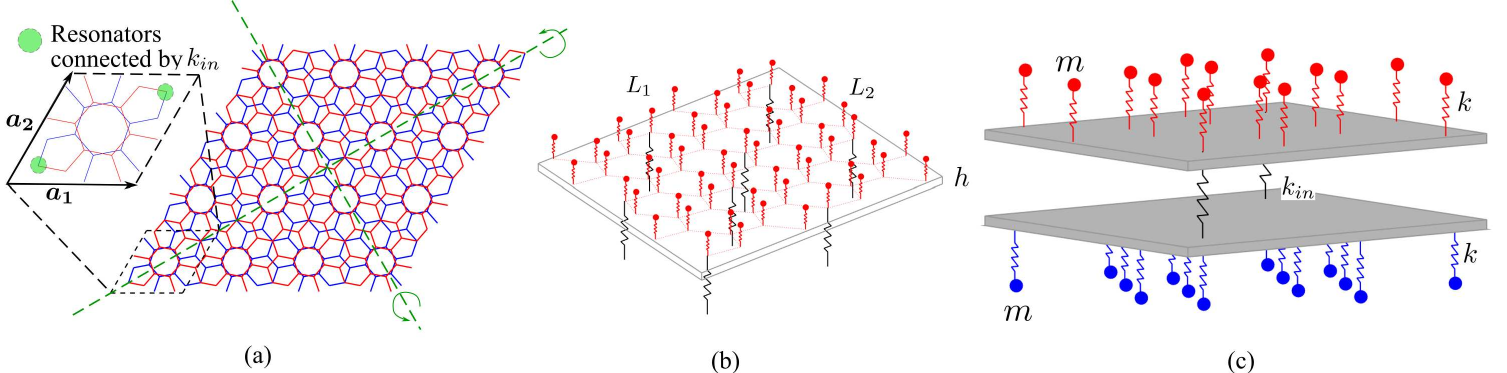}
          \caption{Schematic of bilayer elastic plate lattice. (a) Moir\'{e} lattice with a unit cell in the inset. Green circles indicate locations of coincident resonators in the two layers. Lattice has $C_{2}$ symmetry with respect to both its diagonals (green dash lines). (b) Top plate with resonators (red) in hexagonal lattice configuration. Black springs are at the coincident locations indicated in (a). (c) The moir\'e structure has two plates with resonators, and are coupled by the inter-layer springs, $k_{in}$. 
} 
  \label{fig:sch}
\end{figure}
 
We consider two thin infinite homogeneous and isotropic elastic plates supporting flexural (out-of-plane) vibrations. A set of identical discrete resonators with mass $m$ and stiffness $k$ are connected to each plate in a hexagonal lattice configuration. The resonators are located at the nodes of the hexagons. Let $\bm{r}_{\alpha\beta}$ indicate the position vectors of these resonators in each plate, with the index $\beta$ taking values in $\{t,b\}$ indicating the top and bottom plate, and $\alpha$ is an integer that labels the resonators in each plate. Figure~\ref{fig:sch}b displays a schematic of the top plate with resonators. The bottom plate is rotated at an angle $\phi = 21.78^{\circ}$ with respect to the center of a hexagon so that the resonator locations in the two plates resemble the moir\'{e} lattice as shown in Fig.~\ref{fig:sch}a. Note that the edges of the hexagonal lattice in Figs.~\ref{fig:sch}(a,b) do not have any physical meaning and are shown for clarity. The unit cell of the resulting lattice is indicated by dashed at the bottom left corner in Fig.~\ref{fig:sch}a, along with its expanded view in the inset. Similar to the hexagonal lattice, the lattice vectors of the moir\'e lattice are also at $60^\circ$ to each other. As discussed above, it has 14 nodes in each layer with 2 nodal locations where the top and bottom layers coincide, as indicated by the green circles in the inset. The two plates are coupled by inter-layer springs of stiffness $k_{in}$ at these coinciding locations. Figure~\ref{fig:sch}c displays a schematic of the fully assembled bilayered structure. 

As noted earlier, each layer and thus the infinite lattice has $6$-fold rotation symmetry about an axis through the  unit cell center. In addition, the lattice also has a 2-fold rotation symmetry about both the short and long in-plane diagonals, as indicated by the dashed lines in Fig.~\ref{fig:sch}a. Indeed, when the lattice is rotated by $180^\circ$ about a diagonal, the top plate resonators go to the bottom plate resonators' locations. The resulting structure is thus identical to that prior to rotation. Note that this operation is not equivalent to simply interchanging the top and bottom layers, as the latter will result in a different lattice.

Let us now present the governing equations for elastic waves in this bilayered structure. We assume that the out-of-plane modes are decoupled from the in-plane longitudinal and shear modes. In addition, we assume each resonator has one degree of freedom and can move out-of-plane. The out-of-plane displacement of a resonator located at  $\bm{r}_{\alpha\beta}$ and the mid-plane section of plate $\beta$ are are denoted by $w_{\alpha\beta}$ and $w_{\beta}$, respectively. The dynamics of these thin plates are modeled using the Kirchhoff-Love theory. The  equation of motion of the combined structure having $N$ moir\'e  unit cells is given by~\cite{lopez2022theory,pal2017edge}
 \begin{subequations}
    \begin{align}
    D\nabla^4w_{\beta}+\rho h\ddot{w}_{\beta} &= -\sum_{\alpha = 1}^{14 N} k(w_{\beta}-w_{\alpha\beta})\delta(\bm{x}-\bm{r}_{\alpha\beta})-\sum_{\alpha = 1}^{2 N}k_{in}(w_{\beta}-w_{\beta'})\delta(\bm{x}-\bm{r}_{\alpha\beta}),
    \label{eqn:gov1} \\ 
    m \ddot{w}_{\alpha\beta}&=-k(w_{\alpha\beta}-w_\beta(\bm{r}_{\alpha\beta})).
    \label{eqn:gov2}
\end{align} \label{eqn:GOV}
 \end{subequations} 
Here $\bm{x} = (x,y)$, which denotes the position vector of a point in the plane of the plates, and the gradient operator $\nabla$ in Eqn.~\eqref{eqn:gov1} is with respect to $\bm{x}$. The first term on the right-hand side of Eqn.~\eqref{eqn:gov1} accounts for force due to the resonators, while the last term is for the interaction between the two plates. Subscript $(\beta, \beta')$ in this last term takes values $\{t,b\}$ and $\{b,t\}$ for the top and bottom plates, respectively. The plate bending stiffness is $D=Eh^3/12(1-\nu^2)$, with thickness $h$, Young's modulus $E$, Poisson's ratio $\nu$ and its density is $\rho$. For  $N$ moir\'e unit cells, the number of resonators in each plate and the number of inter-layer springs are $14N$ and $2N$, respectively. The following dimension and properties are chosen for our numerical  calculations: unit cell length $a = 26.5$ mm,  $m=10^{-3}$ kg, $k=10$ kN/m, $k_{in}=2$ kN/m, $h=0.1$ mm, $E=70$ GPa, $\nu=0.33$, $\rho=2700$ kg/$\si{\cubic\meter}$. The material properties correspond to aluminium as the plate material. 

 
\section{Unit cell analysis}\label{sec:sec3}
Having introduced the lattice description and presented the governing equations, we now do a dispersion analysis over its unit cell using the plane wave expansion method. We apply the approach followed in prior works on elastic plates with square and hexagonal array of resonators~\cite{xiao2012flexural,torrent2012acoustic,lopez2022theory}. Then the topological properties of the dispersion bands are determined by computing the fractional corner mode $Q$, which is the elastic analogue of the fractional charge in electronic crystals~\cite{benalcazar2019quantization}. This quantity is used to predict the existence of localized modes at the corners of a finite moir\'e lattice structure. 

\subsection{Dispersion analysis}\label{sec:sec3.1}

\begin{figure}[!htbp]
        \centering
             \subfloat[$k_{in}=0$]{%
              \includegraphics[height=5cm]{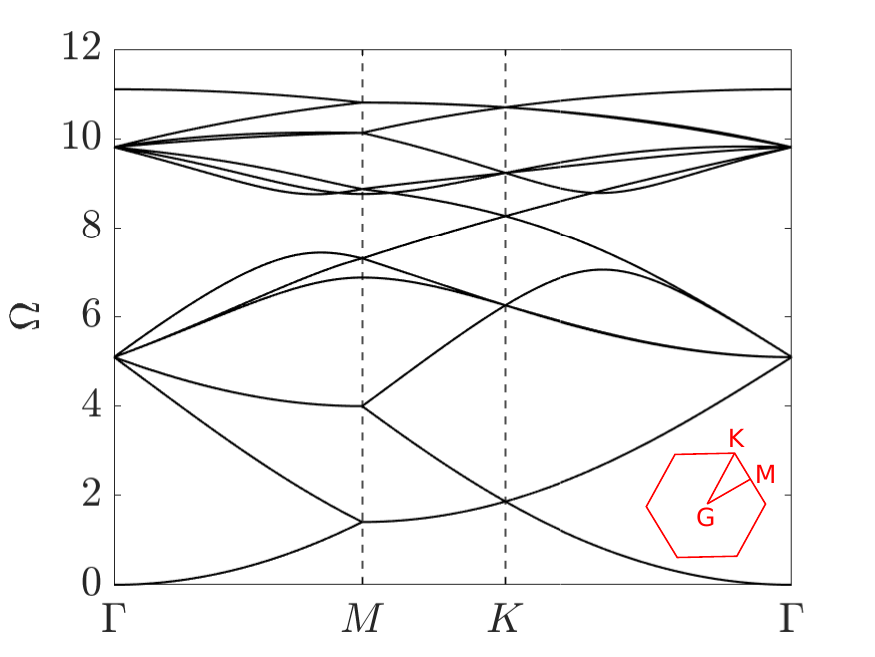}
              \label{fig:disp1}
           } 
             \subfloat[$k_{in}\ne0$]{%
              \includegraphics[height=5cm]{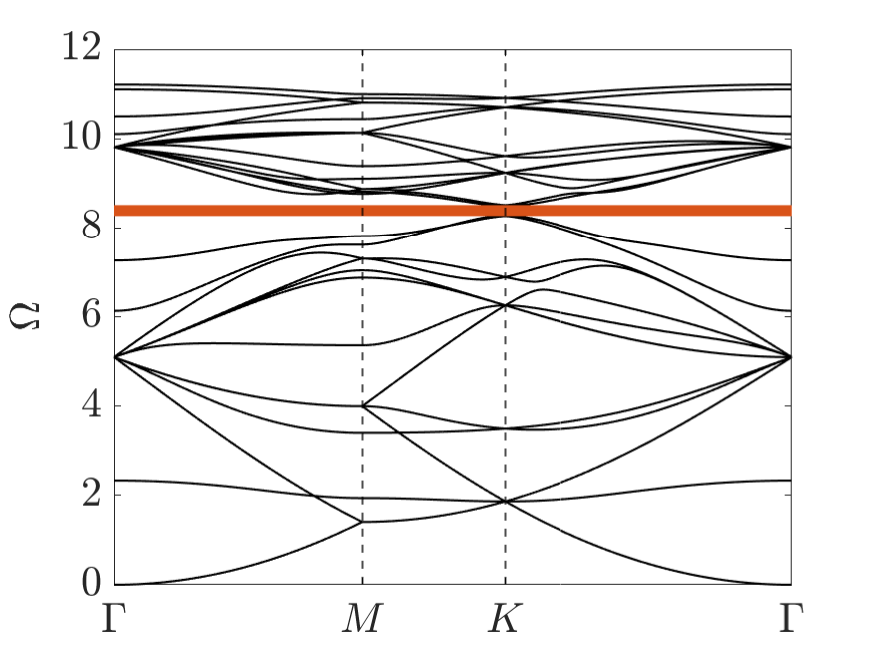}
              \label{fig:disp2}
           } 
           \caption{Dispersion diagram of the unit cell along the IBZ  boundary (a) without inter-layer springs ($k_{in}=0$) and (b) with inter-layer springs ($k_{in}\neq0$). The shaded band in (b) indicates a bandgap opening at the $K$-point.}
           \label{fig:disp}        
\end{figure}

We use Floquet-Bloch theory with the plane wave expansion method to determine waves propagating through the bilayered lattice. For a plane wave propagating with frequency $\omega$ and wave vector $\bkappa$, the displacement field in the plate may be expressed as 
\begin{equation}\label{eqn:Bloch}
 w_\beta(\bm{x},t)
 =e^{i(\omega t+\bm{\kappa} \cdot \bx)} W_{\beta}(\bm{x}). 
\end{equation}
where $W_{\beta}(\bm{x})$ is a periodic function with periodicity of the moir\'{e} unit cell. This periodic function can be expressed using a Fourier series as $\sum_{l_1} \sum_{l_2} W_{l_1 l_2 \beta} e^{i (l_1 \bm{g_1}\cdot \bx + l_2 \bm{g_2}\cdot \bx )}$, with $\bm{g_1,g_2}$ being the reciprocal lattice vectors of the moir\'{e} lattice. They satisfy $\bm{g_i}\cdot  \bm{a_j} = 2\pi \delta_{ij}$ and are given by $\bm{g_1}= 2\pi( 1/a, -1/\sqrt3a)$ and  $\bm{g_2}= 2\pi ( 0 , 2/\sqrt{3}a )$. The summation indices run over all integers, but for computation purposes, we truncate the summations at $T$ terms and use the approximation 
\begin{equation}
 w_\beta(\bm{x},t)
 =e^{i(\omega t+\bm{\kappa.x})} \sum_{\substack{l_1,l_2 = -T}}^T e^{i(l_1\bm{g_1}+l_2\bm{g_2}) \cdot \bm{x}}W_{l_1l_2\beta}.\label{eqn:disp1}  
\end{equation}
Here, $W_{l_1l_2\beta}$ denotes the plane wave coefficient subscripted by integers $l_1,l_2$ and finite number $(2T+1)^2$ of terms are considered. 
The resonator displacement can be expressed as
\begin{equation}
        w_{\alpha\beta}(t)=e^{i(\omega t+\bm{\kappa} \cdot {\bm{r}_{\alpha\beta}})} W_{\overline{\alpha}\beta}.
          \label{eqn:disp2}
\end{equation}
Here the index $\overline{\alpha}$ takes values in $\{1, 2, ..., 14 \}$ and labels the resonators in a reference unit cell, while the index $\alpha$ is an integer that labels resonators in an arbitrary unit cell in the lattice. 

Let us derive the discrete form of the governing equations over the unit cell. Substituting the plate and resonator displacements into Eqn.~\eqref{eqn:gov1}, multiplying by $e^{-i\bm{(k+g').x}}$ and integrating over the unit cell gives an equation for each $W_{l_1 l_2\beta}$. Similarly, substituting the displacements into Eqn.~\eqref{eqn:gov2} gives an equation for each $W_{\bar{\alpha}\beta}$. The detailed derivations are presented in Appendix~\ref{sec:a_dispeqn}. The resulting discretized governing equations are 
\begin{subequations}
\begin{align}
    \omega^2W_{l'_1 l'_2\beta} &= \dfrac{D}{\rho A h}{\lvert \bm{\kappa} + \bm{g'} \rvert^4} W_{l'_1 l'_2\beta}+\frac{k}{\rho A h}\sum_{\substack{\alpha=1}}^{14} e^{-i\bm{g'} \cdot \bm{r}_{\alpha\beta}} \left(\sum_{\substack{l_1,l_2 = -T}}^T e^{i\bm{g} \cdot\bm{r}_{\alpha\beta}} W_{l_1 l_2\beta} - W_{\overline{\alpha}\beta}\right) \nonumber \\ 
    & \qquad\qquad  \qquad  +\frac{k_{in}}{\rho A h}\sum_{\substack{\alpha=1}}^{2} e^{-i\bm{g'}\cdot \bm{r}_{\alpha\beta}}\left( \sum_{\substack{l_1,l_2 = -T}}^T e^{i\bm{g} \cdot \bm{r}_{\alpha\beta}} (W_{l_1 l_2\beta} - W_{l_1 l_2\beta'}) \right) , 
    \label{eqn:disp3} 
    \\ 
\omega^2W_{\overline{\alpha}\beta} &= -\frac{k}{m}\sum_{\substack{l_1,l_2 = -T}}^Te^{i\bm{g}.\bm{r}_{\alpha\beta}}W_{l_1 l_2\beta} + \frac{k}{m}W_{\overline{\alpha}\beta}.
    \label{eqn:disp4}
\end{align}
\end{subequations}
Here $\bm{g}=l_1\bm{g_1}+l_2\bm{g_2}$,  $\bm{g'}=l'_1\bm{g_1}+l'_2\bm{g_2}$ and $A = \sqrt{3} a^2/2$ is the unit cell area. 
Equations~\eqref{eqn:disp3} and~\eqref{eqn:disp4} together constitute an eigenvalue problem and its solution gives the dispersion relation of the unit cell yielding the frequencies $\omega$ at specific wave number, $\bm{\kappa}$. We present results for calculations with $T=10$. Increasing $T$ beyond this value did not result in a noticeable change in the results. Finally, the frequency $\omega$ is expressed in non-dimensional form using the normalization $\Omega=\sqrt{\rho a^4 h/D} \omega $. 

As discussed earlier in Sec.~\ref{sec:plateGovEqn}, the infinite lattice has $6$-fold rotation symmetry about an out of plane axis through the unit cell center and 2-fold rotation symmetry about its in-plane diagonals. Thus its Brillouin zone is a hexagon and its irreducible Brillouin zone (IBZ) comprises of a triangle whose corners are the high symmetry points $\Gamma = (0,0)$, $M=(\pi, \pi/\sqrt{3})$ and $K=(2\pi/3, 2\pi/\sqrt{3})$. Here, we examine the dispersion surfaces along the boundary of the IBZ. 

Figure~\ref{fig:disp} displays the dispersion diagram of a unit cell for points along $\Gamma{\text -}M{\text -}K{\text -}\Gamma$ in the irreducible Brillouin zone (IBZ) for two cases: (a) the plates are uncoupled, $k_{in}=0$ and (b) coupled by inter-layer springs, $k_{in} \neq 0$. The inset in (a) has a schematic of the Brillouin zone, IBZ, along with the high symmetry points. Since the spectrum of $\nabla^4$ operator in Eqn.~\eqref{eqn:gov1} is unbounded, the exact solution has an infinite number of frequencies at each wave vector. There is a huge bandgap above the first 28 dispersion branches in both cases, and we restrict attention to these branches only. For the uncoupled plate case, the dispersion diagram in Fig.~\ref{fig:disp1} has a Dirac cone at $K$ point, consistent with a Dirac cone that arises in a hexagonal lattice. A bandgap opens at that $K$ point when inter-layer springs are added, indicated by the shaded rectangle in Fig.~\ref{fig:disp2}. In addition, two branches become isolated from the remainder of the dispersion curves, consistent with other studies which find isolated flat bands at much smaller moir\'e angles~\cite{lopez2020flat}.

\subsection{Localized mode prediction by computing fractional corner mode $Q$}\label{sec:Qcalc}
We use the dispersion analysis to determine if a finite moir\'e structure has localized modes at its boundary. The bulk edge correspondence principle relates the symmetry and topological properties of the Bloch modes in an infinite lattice to the modes localized on the boundaries of a finite lattice~\cite{benalcazar2017quantized,hasan2010colloquium}. The presence of localized modes can be predicted by computing appropriate topological invariants. Here, we will determine the elastic analog of the fractional corner charge $Q$, which has been introduced to predict and demonstrate localized modes in electronic and photonic media~\cite{benalcazar2017quantized,benalcazar2019quantization,peterson2020fractional}. 

\begin{figure}[!htbp]
        \centering
             \subfloat[$\Gamma$ point ]{%
              \includegraphics[height=4cm]{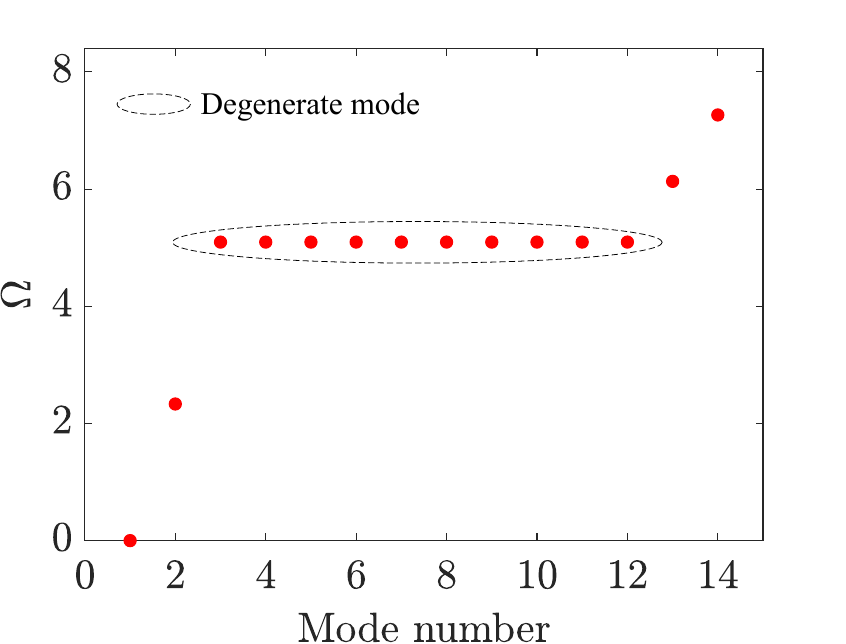}
              \label{fig:gamma}
           } 
             \subfloat[$M$ point]{%
              \includegraphics[height=4cm]{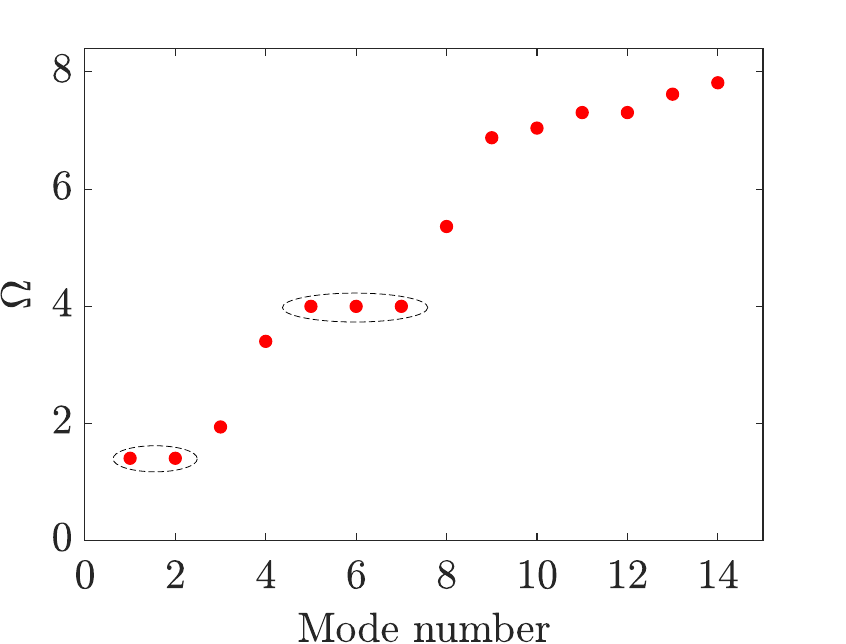}
              \label{fig:M}
           } 
              \subfloat[$K$ point]{%
              \includegraphics[height=4cm]{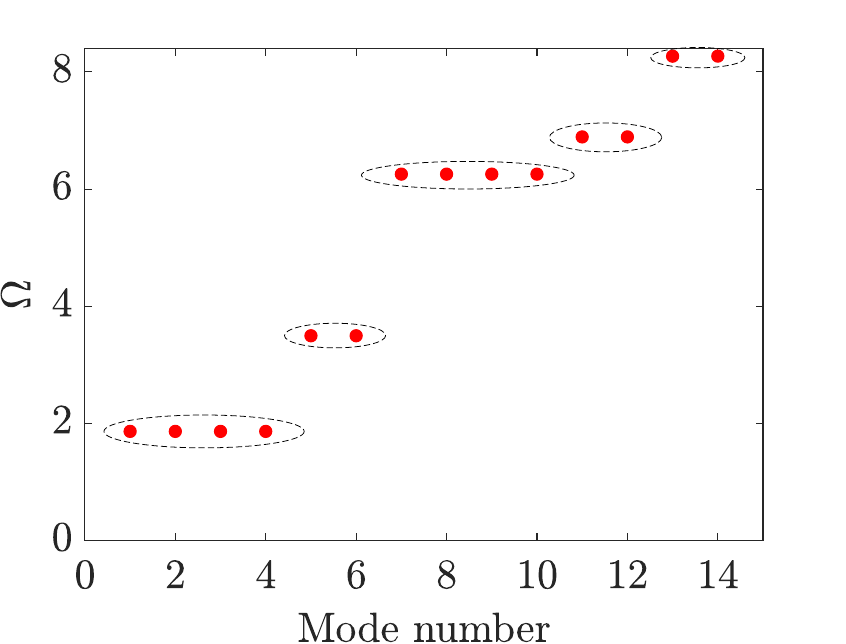}
              \label{fig:K}
           }
           \caption{Frequencies of first $14$ bands below the bandgap at various high symmetry points of the irreducible Brillouin zone.  There are a number of degenerate modes, which require a Gram Schmidt procedure to determine their rotational eigenvalue $\lambda_p$.}
           \label{fig:tiFreq}        
\end{figure}

The fractional corner mode $Q$ is a topological invariant determining the existence of higher order topological mode in the bandgap. This quantity measures the change in rotational symmetry of the Bloch modes as we traverse the dispersion surface. It is expressed in terms of the number of specific rotational symmetry eigenvalues of the Bloch modes at the high symmetry points. All the dispersion branches below the bandgap are considered to compute $Q$. There are 14 bands below the band gap shown in Fig.~\ref{fig:disp2} for the moir\'e lattice. Figure~\ref{fig:tiFreq} displays the distribution of these $14$ frequencies at the various high symmetry points. We note that there are several degenerate sets of frequencies. 

Before computing $Q$, let us discuss how a mode shape at the high symmetry points transforms as the lattice is rotated by an angle about the center of a unit cell. First, let us consider the high symmetry point $K$ and the rotation angle is $2\pi/3$. Under this rotation, the lattice geometry looks identical to that prior to rotation. Let $\bm{R}(\theta)$ the rotation matrix given by 
\begin{equation*}
\bm{R}(\theta)=\begin{pmatrix}
    \cos \theta & -\sin \theta \\
    \sin \theta &  \cos \theta 
\end{pmatrix},
\end{equation*}
and let $\bR_1 = \bR(2\pi/3)$. Recall that the bilayered plate has 6-fold rotation symmetry about a unit cell center and thus remains identical when rotated by $\theta$. Let us indicate a plane wave with  wave vector $\bkappa$ by $w_\beta(\bx,\bkappa)$. It is also a function of $t$ and $\omega$, these are not indicated for brevity. Thus for every plane wave $w_\beta(\bx,\bkappa)$, there is a corresponding plane wave with wave vector $\bR \bkappa$, whose mode shape is $w_\beta(\bR \bx ,\bR \bkappa)$. This condition leads to the relation 
\begin{equation}\label{w_Rot1}
    w_{\beta}\left( \bx , \bkappa \right) = w_{\beta}\left(\bR_1 \bx , \bR_1 \bkappa \right) 
    = w_{\beta}\left(  \bR_1 \bx , \bkappa - \bm{g_1}-\bm{g_2} \right).  
\end{equation}
The second equality in the above equation follows by observing that the wave vector at $K$ satisfies $\bR_1 \bkappa = \bkappa - (\bm{g_1} + \bm{g_2})$, i.e., it translates by $-(\bm{g_1}+\bm{g_2})$ when rotated by $\theta = 2\pi/3$. The Bloch mode shapes at wave vector $\bkappa-\bm{g_1}-\bm{g_2}$ are identical to that at $\bkappa$, as the term $e^{-i (\bm{g_1}+\bm{g_2})\cdot \bx}$ relating them in Eqn.~\eqref{eqn:Bloch} is a periodic function~\cite{bradley2010mathematical}. Each set of corresponding Bloch modes at these two wave vectors may differ by a phase factor $\lambda$ as we continuously traverse the reciprocal lattice~\cite{dresselhaus2007group}. Hence, we have  $w_\beta(\bx,\bkappa -\bm{g_1}-\bm{g_2}) = \lambda w_\beta(\bx, \bkappa)$. Substituting this relation for a point $\bm{R_1 x}$ into the right side of Eqn.~\eqref{w_Rot1}, we see that the displacements at $\bx$ and $\bR_1 \bx$ in a Bloch mode shape at the $K$ point are related by  
\begin{equation}\label{w_Rot}
    w_{\beta}\left( \bx , \bkappa \right) = \lambda w_{\beta} (\bR_1 \bx , \bkappa) .  
\end{equation}
Applying Eqn.~\eqref{w_Rot} successively three times, we get the relation $w_{\beta}(\bx,\bkappa) = \lambda^3 w_{\beta}(\bR_1^3 \bx, \bkappa)$. Noting that $\bR_1^3$ is the identity matrix, we have $\lambda^3 = 1$. Its solutions are 
\begin{equation*}
    \lambda_p = e^{i 2\pi (p-1)/3}, \; p \in \{ 1,2,3 \}.  
\end{equation*}
Thus each mode shape $w_{\beta}(\bx)$ at the high symmetry point $K$ satisfies Eqn.~\eqref{w_Rot} for a specific value of $\lambda_p$. This $\lambda_p$ can thus be viewed as the eigenvalue of the rotational symmetry operator $\bR_1$ for the mode shape. 

Let us now describe the procedure to find the rotational eigenvalue $\lambda_p$ for each Bloch mode at the $K$ point. We project a mode shape into the subspace where a function $u(\bx)$ satisfies  $u(\bR_1 \bx) = \lambda_p u(\bx)$. The projected mode is given by 
\begin{equation*}
w_{\beta}^p(\bm{x}) = \dfrac{1}{3}\left( w_{\beta}(\bm{x}) + \lambda_p^{-1} w_{\beta}(\bm{R_1 x})+ \lambda_p^{-2} w_{\beta}(\bm{R_1^2 x})\right),\; p \in \{ 1,2,3 \} . 
\end{equation*}
By direct substitution, we can verify that any mode shape is decomposed into three parts $w_{\beta}^p$, that satisfy $w_\beta(\bm{x}) = w_{\beta}^1(\bx) + w_{\beta}^2(\bx) +w_{\beta}^3 (\bx)$. If a mode shape has rotational eigenvalue $\lambda_q, q\in \{1,2,3\}$, then the component $w_{\beta}^q$ is non-zero, while the other two projected components are zero.  For example, if a mode satisfies $w_\beta(\bx) = \lambda_1 w_\beta(\bm{R_1} \bx)$, then $w_{\beta}^1(\bx) = w_\beta(\bx)$ and $w_{\beta}^2(\bx)=w_{\beta}^3(\bx)=0$. Thus examining the norms or magnitudes of $w_\beta^p(\bx)$ suffices to identify $\lambda_p$ for a non-degenerate mode. Next, let us discuss how to deal with a set of modes with degenerate frequencies. Here $w^p_\beta(\bx)$, determined by the above equation, can all be non-zero, as $w_\beta(\bx)$ may be a linear combination of mode shapes with distinct $\lambda_p$. To resolve this, we first determine $w_\beta^p(\bx)$ for all the mode shapes, say $n_d$, at a particular degenerate frequency. Then, for each $p$, a Gram-Schmidt procedure is done on the $n_d$ projected modes $w_{\beta}^p(\bx)$. The number of orthogonal modes with non-zero norm gives the  number of independent $w_\beta^p(\bx)$, which is equal to the number of modes with rotational eigenvalue $\lambda_p$ in this set of $n_d$ modes. 

We follow a similar approach to determine the rotational eigenvalues at the other high symmetry points $M$ and $\Gamma$. The wave vector at $M$ satisfies $\bR(\pi) \bkappa = \bkappa - \bm{g_1}-\bm{g_2}$. Again, we note that the lattice looks identical prior to and after rotation by $\theta=\pi$. Using the same steps as for the $K$ point, the corresponding $\lambda_p$ are 
\begin{equation*}
    \lambda_p = e^{i 2\pi (p-1)/2} = (-1)^{p-1}, \; p \in \{ 1,2 \},  
\end{equation*}
and the projected modes are 
\begin{equation*}
w_{\beta}^p(\bm{x}) = \dfrac{1}{2}\left( w_{\beta}(\bm{x}) + (-1)^{p-1} w_{\beta}(\bm{R}(\pi) \bx)  \right),\; p \in \{ 1,2 \} . 
\end{equation*}
Examining the norms of $w^p_\beta$ or using a Gram Schmidt procedure for the sets with degenerate frequencies allows us to determine $\lambda_p$ for each mode. The wave vector at $\Gamma$ point satisfies both $\bR(2\pi/3) \bkappa = \bkappa$ and $\bR(\pi) \bkappa = \bkappa$. For each mode at $\Gamma$, we can thus determine the rotational eigenvalues $\lambda_p$ for  rotations by $2\pi/3$ and $\pi$. The mode shapes $w^p_\beta$ corresponding to each $\lambda_p$ for all the high symmetry points are presented in Appendix~\ref{sec:a_ti}, see Figs.~\ref{fig:ti1}-~\ref{fig:ti4}. 


The fractional corner mode $Q$, analogous to its electronic counterpart, is given by~\cite{liu2021higher}
\begin{equation}
    Q=\dfrac{[M_1^{(2)}]}{4}+\dfrac{[K_1^{(3)}]}{6} \mod 1 . 
    \label{eqn:ti1}
\end{equation}
Here, $[M_1^{(2)}]$ is the difference between the number of modes at $M$ and $\Gamma$ points that have $\lambda_p=1$ under rotation by $\theta=\pi$.  Similarly, $[K_1^{(3)}]$ denotes the difference between the number of mode shapes at $K$ and $\Gamma$ points  with $\lambda_p=1$ under rotation by $\theta=2\pi/3$. Counting the number of mode shapes with $p=1$ at the high symmetry points below the bandgap, we have from Eqn.~\eqref{eqn:ti1} 
\begin{equation}
    Q=\dfrac{(6-8)}{4}+\dfrac{(4-6)}{6} =\dfrac{-5}{6} =\dfrac{1}{6} \mod 1 . 
    \label{eqn:Q}
\end{equation}
A non-zero value of $Q$ in Eqn.~\eqref{eqn:Q} confirms the non-trivial topological  nature of the bandgap, which in turn, implies the existence of corner localized modes in a finite structure. 

We apply the framework established by Hughes and coworkers~\cite{benalcazar2019quantization} in the context of electronic waves and charges to predict the location of localized modes. This framework allows us to express the stiffness matrix of any  $C_6$ symmetric structure as a direct sum of copies of the stiffness matrices of primitive generator lattices. The topological invariants, like $Q$, are a sum of the corresponding $Q$ values of these primitive generators. We consider two primitive generators: $h_{4b}$, $h_{3c}$ that have nontrivial topological properties. Here, a lattice with notation $h_{mW}$ has  $m$ bands below the bandgap and a Wannier center at location $W$~\cite{benalcazar2019quantization}. The lattice schematics, unit cell and dispersion diagrams for these two lattices are presented in Appendix~\ref{sec:a_prim}. 

We computed the fractional corner modes for these primitive generator lattices by considering both 6-fold and 3-fold rotation symmetry. The values are $Q_6=2/3,\;Q_3=1/3$ for the $h4b$ lattice and  $Q_6=1/2,\; Q_3=0$ for the $h3c$ lattice. Here, the subscripts of $Q$ indicate the rotation symmetry of the finite structure. Thus $Q_6$ and $Q_3$ determine localized modes at $120^\circ$ and $60^\circ$ corners, respectively. These values show that the $h4b$ lattice has a localized mode at $60^\circ$ corner, while both lattices have at $120^\circ$ corner. 
Noting that $Q$ of the moir\'e lattice may be expressed as $Q = 1/6 = 2/3 + 1/2 \;(\mod 1)$, we infer that the moir\'{e} lattice is equivalent to stacking a copy of each of these these two primitive generators, along with copies of a lattice ($t$) that has trivial topological properties. In other words, the stiffness matrix of the moir\'e lattice, expressed in the basis of the first 28 dispersion bands, is equivalent to the direct sum $h4b \oplus h3c \oplus 7t$. This direct sum, along with the $Q$ values of the primitive generators, indicates the existence of corner localized modes at both $120^\circ$ and $60^\circ$  corners in our moir\'{e} structure.  Indeed, the latter case of $60^\circ$ corner localized mode is inferred by noting that the $Q_3$ value of our moir\'e lattice is $1/3+0 = 1/3$.

\section{Numerical results of finite plate}\label{sec:numer}
In this section, the predictions of corner localized modes in Sec.~\ref{sec:Qcalc} are verified by determining the mode shapes and frequency response under external excitation on a finite plate. We show that these localized modes are excited even when an external force is applied far from the corner. 

\subsection{Bulk and localized mode shapes}

\begin{figure}[!htbp]
        \centering
     \includegraphics[height=8.4cm]{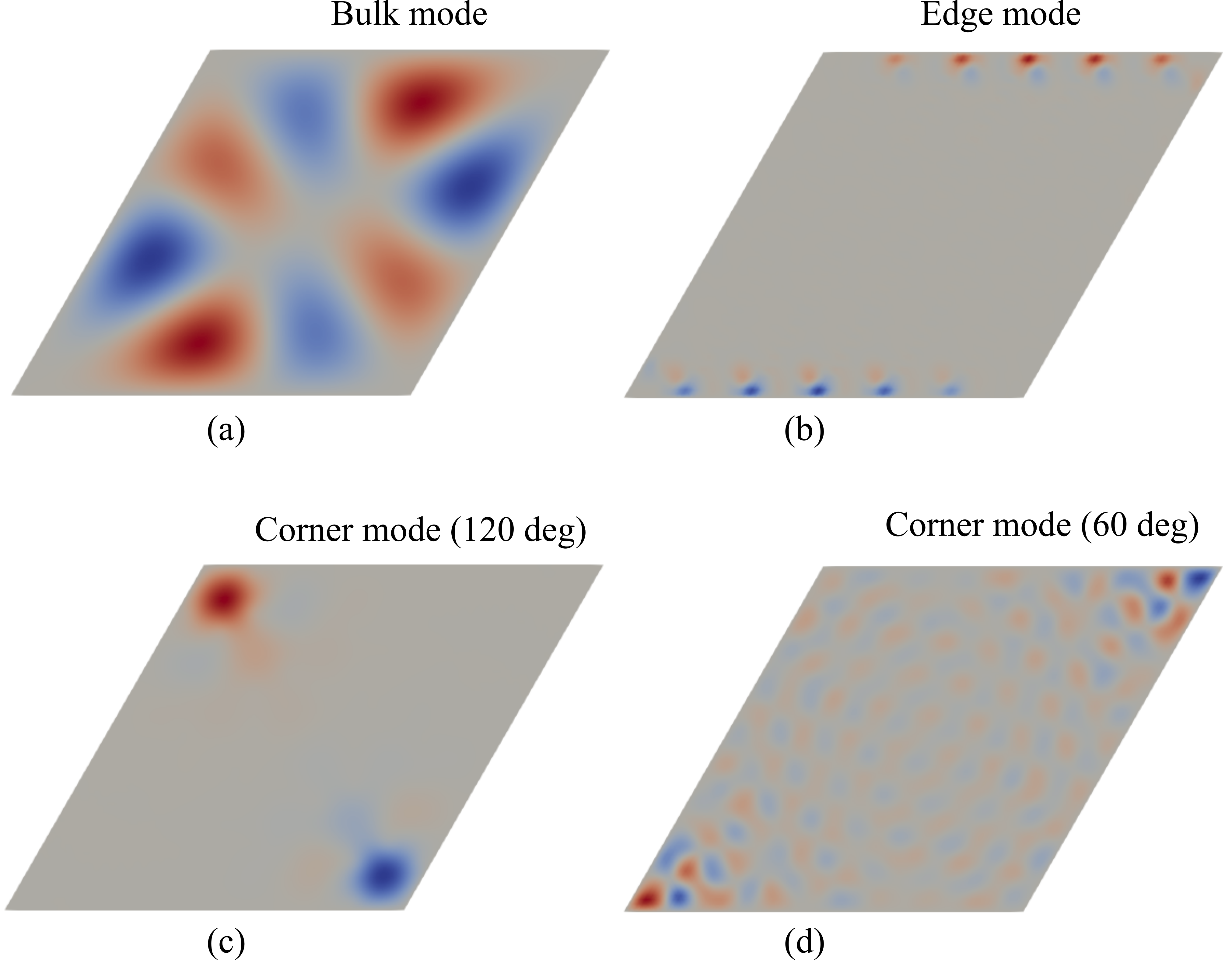}
          \caption{Mode shapes of a finite plate with simply-supported boundary conditions. Displacement contours of top plate for (a) a typical bulk mode ($\Omega=0.70$), (b) an edge mode ($\Omega=13.71$), corner modes at (c)  $120^{\circ}$ corner ($\Omega=2.34$) and (d) $60^{\circ}$ corner ($\Omega=8.19$).  
} 
  \label{fig:ms}
\end{figure}

We consider a finite plate of  $n_1\times n_2 = 6\times6$ moir\'{e} unit cells along $\bm{a_1}$, $\bm{a_2}$ directions. The sides of the plate are of lengths $L_1$ and $L_2$, both equal to $6a$. The four sides of the plate are simply supported, implying zero displacement and zero bending moment about an axis along the boundary. At each boundary point, these conditions may be expressed as~\cite{timoshenko1959theory} 
\begin{equation*}
    w_{\beta} = 0, \quad M_{\eta} = \dfrac{\partial^2 w_{\beta} }{\partial \eta^2} + \nu \dfrac{\partial^2 w_{\beta} }{\partial \tau^2} = 0, 
\end{equation*}
with $\eta$ and $\tau$ being coordinates normal to and along the boundary. 

We introduce and work with a coordinate system whose axes $(x_1,x_2)$ are aligned with the lattice vectors of the moir\'e lattice. The boundary conditions and solution basis functions are conveniently expressed in this coordinate system. To determine the governing equations in this coordinate system, let us determine its relation with the Cartesian coordinate system having axes $(x,y)$. Let us consider an arbitrary point with position vector $\bx$ in the two coordinate systems. It is given by $\bm{x}$ = $x\bm{e_x}$ + $y\bm{e_y} = x_1\bm{e_1}$ + $x_2\bm{e_2}$, with $(\bm{e_x},\bm{e_y})$ and $(\bm{e_1},\bm{e_2})$ being unit vectors in the two coordinate systems. Taking dot products with $\bm{e_x}$ and $\bm{e_y}$ gives the relations
$x = x_1 +  x_2/2$ and $y = \sqrt{3} x_2/2$. They can be inverted to get $x_1 = x-y/\sqrt{3}$ and $x_2 = 2y/\sqrt{3}$.

The boundary conditions in the new coordinate system become
\begin{gather*}
    w_\beta(x_1=0, x_2)=w_\beta(x_1=L_1, x_2)=w_\beta(x_1, x_2=0)=w_\beta(x_1, x_2=L_2) = 0 ,\\ 
     \pdv[2]{w_\beta}{x_1}\bigg|_{(x_1=0)}=\pdv[2]{w_\beta}{x_1}\bigg|_{(x_1=L_1)}=\pdv[2]{w_\beta}{x_2}\bigg|_{(x_2=0)}=\pdv[2]{w_\beta}{x_2}\bigg|_{(x_2=L_2)}=0 . 
\end{gather*}
The plate displacement, $w_{\beta}(\bm{x},t)$ is approximated using a set of harmonic basis functions as 
\begin{equation}
 w_{\beta}(\bm{x},t)=e^{i\omega t} \sum_{\substack{p = 1}}^{N_1} \sum_{\substack{q = 1}}^{N_2} \sin{\frac{p \pi x_1}{L_1} \sin{\frac{q \pi x_2}{L_2}}} W_{pq\beta} .
  \label{eqn:ms1}
\end{equation}
Note that these basis functions satisfy all the above boundary conditions. Similarly, the resonator displacements, $w_{\alpha\beta}(t)$ can be written as 
\begin{equation}
        w_{\alpha\beta}(t)=e^{i\omega t} W_{\alpha\beta},
         \label{eqn:ms2}
\end{equation}
with the index $\alpha$ ranging from $1$ to $14 \times n_1 \times n_2$. 

Let us now derive the discrete approximations of the governing equations for vibration at frequency $\omega$. Substituting the above displacements into Eqn.~\eqref{eqn:gov1}, multiplying by $\sin(p' \pi x_1 / L_1) \sin( q' \pi x_2 /L_2)$ and integrating over the finite plate leads to an equation for each basis function. 
Similarly, substituting the displacements into Eqn.~\eqref{eqn:gov2} gives an equation for each resonator displacement amplitude $W_{{\alpha}\beta}$. The detailed derivations are presented in Appendix~\ref{sec:a_finite}. The discretized governing equations thus obtained are 
\begin{subequations}
\begin{align}
\omega^2 W_{p' q' \beta}&=\frac{16 \pi^4 D }{9 \rho h L_1^4L_2^4 }\left(p'^4L_2^4+ 3 p'^2q'^2 L_1^2L_2^2+q'^4L_1^4\right)  W_{p'q'\beta} \nonumber\\ 
& \quad +\frac{32p' q' D }{9 \rho h }\sum_{\substack{p = 1 \\ p \ne p^\prime}}^{N_1} \sum_{\substack{q = 1 \\ q \ne q^\prime}}^{N_2} \frac{p q}{L_1 L_2}\left\{(\frac{p \pi}{L_1})^2 + (\frac{q \pi}{L_2})^2\right\} 
\left[\frac{(-1)^{p+p'} -1}{p^2-p'^2}\right]
\left[\frac{(-1)^{q+q'} -1}{q^2-q'^2}\right] W_{pq\beta} \nonumber\\
&\quad +\frac{8k}{\sqrt{3}\rho h L_1 L_2 } \sum_{\alpha = 1}^{14 n_1 n_2}\sin{\frac{p' \pi r_{\alpha\beta1}}{L_1} \sin{\frac{q' \pi r_{\alpha\beta2} }{L_2}}}\left[\sum_{p=1}^{N_1} \sum_{q=1}^{N_2}
\sin{\frac{p \pi r_{\alpha\beta1} }{L_1} \sin{\frac{q \pi r_{\alpha\beta2} }{L_2}}}W_{pq\beta} -W_{\alpha\beta}\right] \nonumber\\ 
&\quad +\frac{8k_{in}}{\sqrt{3}\rho h L_1 L_2} \sum_{\alpha=1}^{2 n_1 n_2}\sin{\frac{p' \pi r_{\alpha 1}}{L_1} \sin{\frac{q' \pi r_{\alpha 2}}{L_2}}}\left[\sum_{p=1}^{N_1} \sum_{q=1}^{N_2} \sin{\frac{p \pi r_{\alpha 1}}{L_1} \sin{\frac{q \pi r_{\alpha 2}}{L_2}}}(W_{pq\beta} -W_{pq\beta'})\right] ,
   \label{eqn:ms3} \\ 
    \omega^2 W_{\alpha\beta}&=\dfrac{k}{m}W_{\alpha \beta}-\dfrac{k}{m}\sum_{p=1}^{N_1} \sum_{q=1}^{N_2}  \sin{\frac{p \pi r_{\alpha \beta 1}}{L_1} \sin{\frac{q \pi r_{\alpha \beta 2}}{L_2}}} W_{pq\beta}.
    \label{eqn:ms4}
\end{align}
\end{subequations}
Here $\br_{\alpha \beta} = r_{\alpha\beta 1} \be_1 + r_{\alpha\beta 2} \be_2$ and $\br_{\alpha} = r_{\alpha 1} \be_1 + r_{\alpha 2} \be_2$ are the position vectors of resonators and inter-layer springs expressed in the $(x_1,x_2)$ coordinate system. 
Here, Eqn.~\eqref{eqn:ms3} and Eqn.~\eqref{eqn:ms4} together constitute an eigenvalue problem of the form  $\omega^2\bm{v}=\bm{Kv}$, with $\bm{v}=[\bm{W_{pqt}}; \bm{W_{pqb}}; \bm{W_{\alpha t}};\bm{W_{\alpha b}}]$  being the vector whose components are coefficients of basis functions for both the plate and resonator displacements. $\bm{K}$ is the stiffness matrix containing the right-hand side terms in Eqn.~\eqref{eqn:ms3} and Eqn.~\eqref{eqn:ms4}. The solution of the eigenvalue problem provides the mode shapes at the corresponding frequencies, $\omega$. Each mode shape has 4 parts: displacement fields of the top and bottom plates $w_\beta$, and the vector of resonator displacements $W_{\alpha\beta}$ in each plate. 

Let us remark on the relation between the displacement fields in the plates based on symmetry considerations. Note that the finite bilayer structure also has $C_2$ symmetry about each of its diagonals, similar to the infinite lattice (see Fig.~\ref{fig:sch}a). The mode shapes of the finite plate are thus eigenvectors of this symmetry operator. Since the $C_2$ rotation operator has eigenvalues $\lambda=\pm 1$, each mode shape remains the same or changes sign under a  rotation by $\pi$ along a diagonal. We observe that this symmetry operation is equivalent to reflecting each plate in its plane about a diagonal, followed by interchanging the two plates. Thus, an equivalent way to express the above symmetry condition is the following: for each mode shape, if the top plate displacement field is reflected about a diagonal, it will be same $(\lambda=+1)$ or negative $(\lambda=-1)$ of the bottom plate displacement field. Note that the $\lambda$ values can be distinct when reflected about the short and long diagonals for a mode shape.

The results are reported for calculations with $N_1 = N_2=50$ terms. We also did calculations with $N_1 = N_2=60$, and did not observe a noticeable difference in the mode shapes. Displacement contours at the top plate are illustrated in Fig.~\ref{fig:ms} for a few representative mode shapes. The bottom plate displacement field, top and bottom resonator displacements for these mode shapes are presented in Appendix~\ref{sec:a_ms}, see Figs.~\ref{fig:a_ms1} and~\ref{fig:a_ms2}. We find that the resonator displacements are in phase with their plate displacements for all of these modes. As discussed above, a mode shape may change sign or remain unchanged under rotation by $\pi$ about a diagonal. This relation may be determined by examining the displacement fields of top and bottom plates. They are tabulated below for each mode in Fig.~\ref{fig:ms} and for each diagonal rotation axis. The modes that remain identical and that change sign under rotation are labeled even and odd, respectively.  
\begin{table}[h]
\begin{center}
\begin{tabular}{ |c| c c c c|}
\hline
 - & Bulk & Edge & Corner($120^{\circ}$) & Corner($60^{\circ}$)\\ 
 \hline
 Short diagonal & odd & even & odd & odd \\  
 Long diagonal & odd & odd & even & even \\
 \hline
\end{tabular}
\end{center}
    \caption{Symmetry property of the modes in Fig.~\ref{fig:ms} under $180^\circ$ rotation about its short and long diagonals.}
    \label{tab:table1}
\end{table}

The edge localized mode in Fig.~\ref{fig:ms}b has a counterpart at the same frequency, that is localized at the other edges. The mode shape of this counterpart is included in Appendix~\ref{sec:a_ms}. These edge modes lie in the bandgap above the first 28 dispersion branches. They do not have a topological origin and may become bulk modes when boundary conditions or material properties are varied. In contrast, the corner localized modes shown in Fig.~\ref{fig:ms}(c-d) arise due to the symmetry and topological properties of dispersion bands. The mode localized at the $120^{\circ}$ lies in the bulk band frequency, while the $60^{\circ}$ corner localized mode lies in the bandgap. These localized modes verify the prediction of higher order topological mode at both corners as discussed in Sec.~\ref{sec:Qcalc}.

\subsection{Frequency response  under harmonic excitation}
\begin{figure}[!tb]
        \centering
     \includegraphics[height=3.8cm]{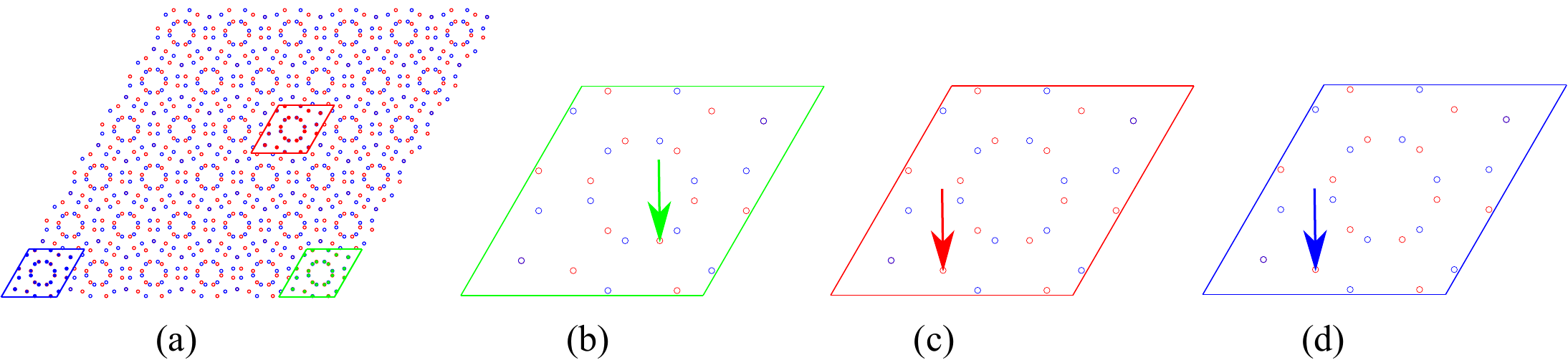}
          \caption{Schematic of the finite lattice showing different excitation and response locations. (a) Responses are evaluated over the unit cell indicated by parallelogram at $120^\circ$ corner, interior and $60^\circ$ corner. Excitations are given at a resonator in the top plate indicated by arrow in (b) $120^\circ$ corner, (c) interior and (d) $60^\circ$ corner.
} 
  \label{fig:frfSch}
\end{figure}

Finally, let us analyze the effect of these topological corner localized modes on the steady-state dynamic response under external excitation. To this end, we determine the frequency response by applying  a harmonic force and measuring the steady state response at various locations (see Fig.~\ref{fig:frfSch}) in the finite lattice. An excitation $f e^{i\omega t}$ is applied at one resonator in the top plate, indicated by an arrow in Figs.~\ref{fig:frfSch}(b-d). The equations for solving the frequency response is $\omega^2\bm{v}-\bm{Kv}=\bm{f}$, where $\bm{f}$ is the external force vector. It takes $1$ associated to the excitation point and rest values are $0$. The considered frequency spacing in the calculation is $\Delta\omega = 1.0\times 10^{-2}$ rad/s, or in non-dimensional units, $\Delta\Omega = \num{4.50e-5}$. For each case, responses are also observed at the $120^{\circ}$ corner, interior and $60^{\circ}$ corner region over a unit cell indicated by parallelogram in Fig.~\ref{fig:frfSch}a. The response is computed using the expression 
\begin{equation}
    \frac{|u|}{f} = \sqrt{\sum_{\beta=
    \{t,b\}}\sum_{\alpha=1}^{14}
    \left|w_{\alpha \beta}\right|^2} .
    \label{eqn:frf}
\end{equation}

To illustrate the effect of corner localized modes in Fig.~\ref{fig:ms}(c-d) on the frequency response function of a finite plate, we excite it in a range of frequencies around these natural frequencies. Recall that the $120^\circ$ and $60^\circ$ corner localized mode frequencies are 2.34 and 8.19, respectively. At each frequency, we excite the lattice at 3 locations: at a $60^\circ$ and a $120^\circ$ corner, and in the interior, and determine the response of the unit cells at these locations using Eqn.~\eqref{eqn:frf}. These locations are indicated in Fig.~\ref{fig:frfSch}. 
\begin{figure}[!htbp]
        \centering
     \includegraphics[height=10cm]{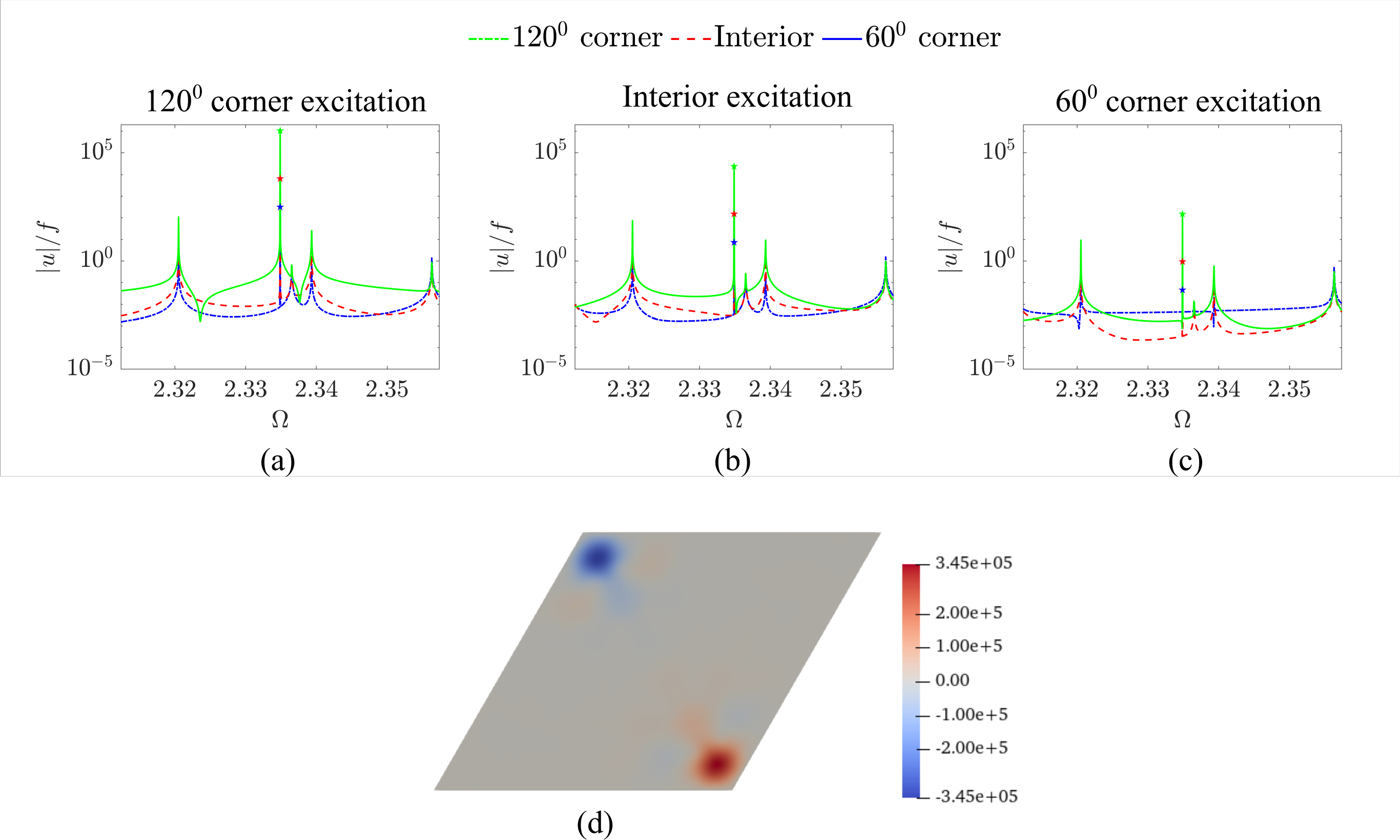}
          \caption{frequency response and displacement contour near the $120^\circ$ corner localized frequency. Excitation is given at (a) $120^\circ$ corner, (b) interior and (c) $60^\circ$ corner and response are also shown at $120^\circ$ corner, interior and $60^\circ$ corner. (d) Top plate displacement contour for $120^\circ$ corner excitation at frequency 2.34. 
} 
  \label{fig:frf1}
\end{figure}

 \begin{figure}[!htbp]
        \centering
     \includegraphics[height=10cm]{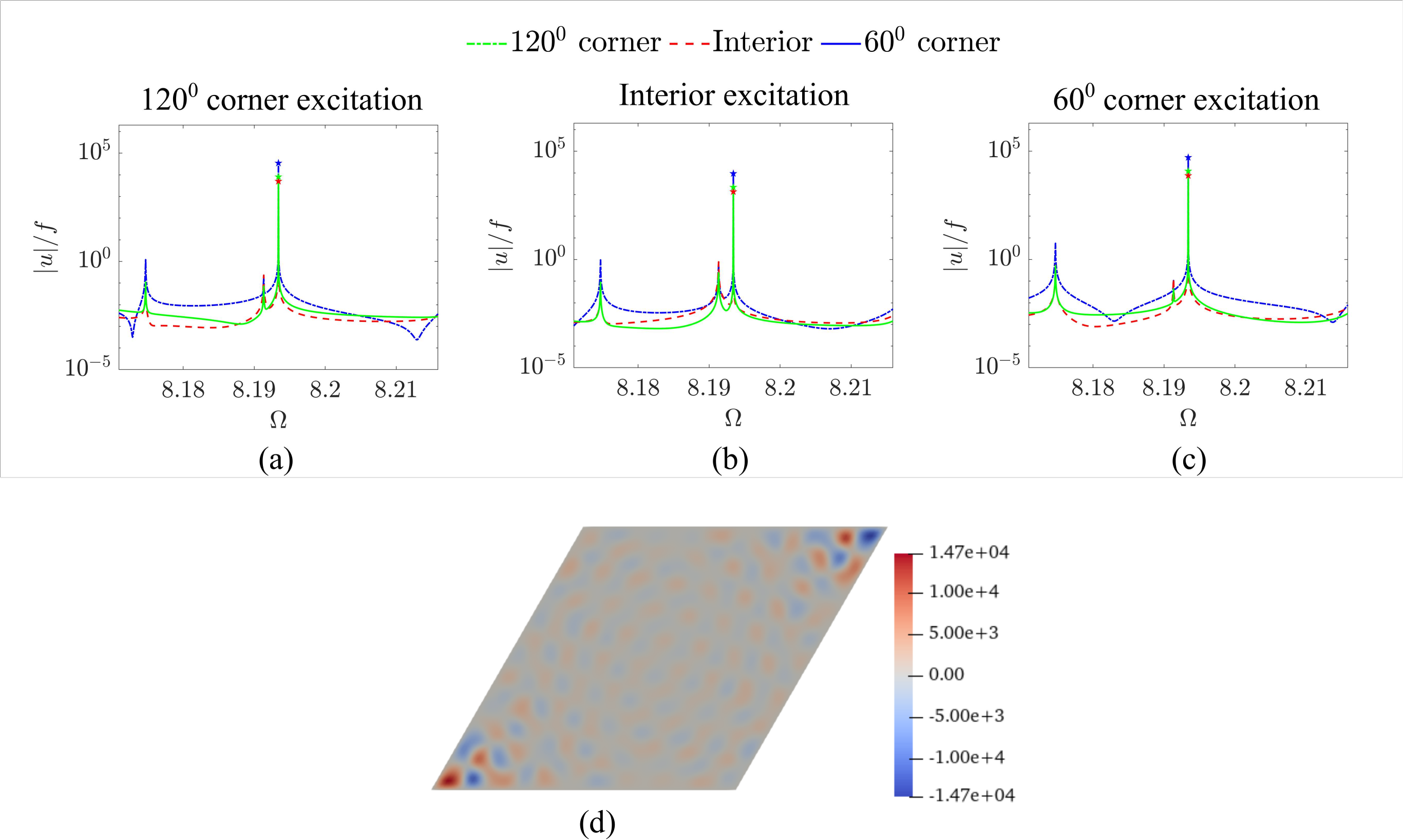}
          \caption{frequency response and displacement contour near the $60^\circ$ corner localized frequency. Excitation is given at (a) $120^\circ$ corner, (b) interior and (c) $60^\circ$ corner and response are also shown at $120^\circ$ corner, interior and $60^\circ$ corner. (d) Top plate displacement contour for $60^\circ$ corner excitation at frequency 8.19. 
} 
  \label{fig:frf2}
\end{figure}
Figure~\ref{fig:frf1}(a-c) displays the frequency response near $\Omega=2.34$, with each sub-figure for a different excitation location. For the $120^\circ$ corner excitation, Fig.~\ref{fig:frf1}a displays the response at various locations in the finite plate indicated in Fig.~\ref{fig:frfSch}a. The peak responses at all locations happen  at frequency 2.34, as indicated by a ``star" in the figure. The response of the $120^\circ$ corner unit cell is higher than at other locations, since it is close to the excitation point. Similarly, Fig.~\ref{fig:frf1}(b-c) displays the response for excitations at the interior and $60^\circ$ corner locations. Even when the excitation is far from the $120^\circ$ corner, the peak response at the localized mode frequency is the highest at this corner. This peak response shows that the corner localized mode gets excited regardless of the excitation location in the plate. In contrast, away from the localized mode frequency, we note that the response is higher close to the excitation location. See for example, the response to $60^\circ$ corner excitation in Fig.~\ref{fig:frf1}c. 
Figure~\ref{fig:frf1}d displays the displacement contours of the top plate for an excitation $\Omega=2.34$ at $120^\circ$ corner, which is similar to the mode shape in Fig.~\ref{fig:ms}c. The displacement contours for excitation at interior and $60^\circ$ corner locations have similar profile, but with lower peak magnitudes of ${7.96}\times 10^{3}$ and ${4.95}\times 10^{1}$, respectively.  

Similarly, Fig.~\ref{fig:frf2}(a-c) displays the frequency response around $\Omega=8.19$ for various excitation locations. Again, the peak response happens at the localized mode frequency and the $60^\circ$ corner has the highest displacement magnitude $|u|/f$, regardless of the excitation location. Figure~\ref{fig:frf2}d illustrates the displacement contour of top plate for excitation at $60^\circ$ corner, confirming that the steady-state response is localized at  the  $60^\circ$ corner. The displacement contours for excitation at the $120^\circ$ corner and interior have a similar profile, but with different maximum magnitudes of ${9.87}\times 10^{3}$ and ${2.66}\times 10^{3}$, respectively. These calculations verify the presence of corner localized modes at both corners of the finite moir\'e plates. 

\section{Conclusions}\label{sec:conc}
We investigated corner localized modes that arise due to higher order topology in moir\'{e} lattices of bilayer elastic plates. Each plate has a hexagonal array of resonators and one of the plates is rotated an angle ($21.78^\circ$) which results in a periodic moir\'{e} lattice with the smallest area. The resulting structure opens a band gap when inter-layer springs are added. The fractional corner mode $Q$ is found to be $1/6$ for dispersion bands below the bandgap. The non-zero value of $Q$ indicates the non-trivial topological nature of the bandgap and predicts the existence of localized modes at all corners in the finite structure. Modal analysis on a finite plate showed the existence of these corner localized modes at both $60^\circ$ and $120^\circ$ corners. The first one lies in the bulk band frequency and the later one lies in the bandgap frequency. Finally, the frequency response under external excitation at various locations shows mode localization at these frequencies, consistent with the theoretical predictions. The considered continuous elastic moir\'e lattice structures open opportunities for seeking novel wave phenomena with potential applications in tunable energy localization, vibration isolation, and energy harvesting.

\begin{acknowledgments}
This work was supported by the U.S. National Science Foundation under Award No. 2238072. 
\end{acknowledgments}

\appendix

\section{Derivation of discrete governing equations}
The detailed derivation of the discrete form of the governing equations in terms of the Fourier coefficients $W_{l_1 l_2 \beta}$ are presented for dispersion analysis and $W_{pq\beta}$ for finite plate analysis. 

\subsection{Dispersion analysis}\label{sec:a_dispeqn}
We start by substituting the displacements in Eqns.~\eqref{eqn:disp1} and~\eqref{eqn:disp2} into the governing equation for the plate~\eqref{eqn:gov1}, which leads to  
\begin{multline}
    (D{\lvert \bm{\kappa} + \bm{g} \rvert^4}-\rho h \omega^2) \sum_{\substack{l_1,l_2 = -T}}^T e^{i\bm{(k+g).x}}W_{l_1l_2\beta} =
 -k\sum_{\alpha = 1}^{14 N} \left[ \sum_{\substack{l_1,l_2 = -T}}^T e^{i\bm{(k+g).x}}W_{l_1l_2\beta} - e^{i\bm{\kappa.{\bm{r}_{\alpha\beta}}}}W_{\overline{\alpha}\beta}\right] \delta(\bm{x}-\bm{r}_{\alpha\beta}) \\
     -k_{in}\sum_{\alpha = 1}^{2 N} \left[ \sum_{\substack{l_1,l_2 = -T}}^T e^{i\bm{(k+g).x}}(W_{l_1l_2\beta}-W_{l_1l_2\beta'})\right] \delta(\bm{x}-\bm{r}_{\alpha\beta}) .
\end{multline}
We work in the $(x_1,x_2)$ coordinate system, whose unit vectors are aligned with the moir\'e lattice vectors $(\bm{a_1},\bm{a_2})$. It is related to the Cartesian coordinate system by $x_1 = x-y/\sqrt{3}$ and $x_2 = 2y/\sqrt{3}$. 
Multiplying by $e^{-i\bm{(k+g').x}}$, rearranging and integrating over a unit cell $\mathcal{D}$ gives
\begin{multline}
    \int_{\mathcal{D}}\sum_{\substack{l_1,l_2 = -T}}^T(D{\lvert \bm{\kappa} + \bm{g} \rvert^4}-\rho h \omega^2) e^{i\bm{(g-g').x}}W_{l_1l_2\beta} dx_1 \wedge dx_2 
    \\
    = -k\int_{\mathcal{D}}\sum_{\alpha = 1}^{14 N} \left[e^{-i\bm{g'.x}}\left\{\sum_{\substack{l_1,l_2 = -T}}^T e^{i\bm{g.x}}W_{l_1l_2\beta} - e^{i\bm{\kappa.({r}_{\alpha\beta}-x)}} W_{\overline{\alpha}\beta}\right\}\right] \delta(\bm{x}-\bm{r}_{\alpha\beta})dx_1 \wedge dx_2
    \\
    -k_{in}\int_{\mathcal{D}}\sum_{\alpha = 1}^{2 N} \left[e^{-i\bm{g'.x}}\left\{\sum_{\substack{l_1,l_2 = -T}}^T e^{i\bm{g.x}}(W_{l_1l_2\beta} - W_{l_1l_2\beta'}) \right\}\right] \delta(\bm{x}-\bm{r}_{\alpha\beta})dx_1 \wedge dx_2 . 
\end{multline}
Note that $dx_1 \wedge dx_2 = (\sqrt{3}/2) dx_1 dx_2$ is the area of an infinitesimal parallelogram in $\mathcal{D}$. Using orthogonality of the functions $e^{i\bm{g} \cdot \bm{x}}$, we get 
\begin{multline}
    (D{\lvert \bm{\kappa} + \bm{g'} \rvert^4}-\rho h \omega^2)AW_{l_1'l_2'\beta} \nonumber= -k\sum_{\alpha = 1}^{14 N} \left[e^{-i\bm{g'.\bm{r}_{\alpha\beta}}}\left\{\sum_{\substack{l_1,l_2 = -T}}^T e^{i\bm{g.\bm{r}_{\alpha\beta}}}W_{l_1l_2\beta} - W_{\overline{\alpha}\beta}\right\}\right] \\
     -k_{in}\sum_{\alpha = 1}^{2 N} \left[e^{-i\bm{g'.\bm{r}_{\alpha\beta}}}\left\{\sum_{\substack{l_1,l_2 = -T}}^T e^{i\bm{g.\bm{r}_{\alpha\beta}}}(W_{l_1l_2\beta} - W_{l_1l_2\beta'}) \right\}\right] . \nonumber
\end{multline}
Here $A = \sqrt{3} a^2/2$ is the area of the moire unit cell. Dividing both sides of the above equation by $\rho A h$ and rearranging gives Eqn.~\eqref{eqn:disp3}.

For the resonator, substituting the displacements in Eqns.~\eqref{eqn:disp1} and~\eqref{eqn:disp2} into their governing equations~\eqref{eqn:gov2} gives
\begin{equation}
-m\omega^2e^{i{\bm{\kappa.r_{\alpha\beta}}}}W_{\overline{\alpha}\beta} = -k e^{i{\bm{\kappa.r_{\alpha\beta}}}} W_{\overline{\alpha}\beta}+k\sum_{\substack{l_1,l_2 = -T}}^Te^{i\bm{(k+g)}.\bm{r_{\alpha\beta}}}W_{l_1l_2\beta} . \nonumber  
\end{equation}
Multiplying by $-e^{-i{\bm{\kappa.r_{\alpha\beta}}}}/m$ gives Eqn.~\eqref{eqn:disp4}. 

\subsection{Derivation for finite plate frequencies and mode shapes}\label{sec:a_finite}
Let us derive the discrete equations that are used to determine the mode shapes and frequency response of a finite plate. Substituting the assumed displacement fields in Eqns.~\eqref{eqn:ms1} and~\eqref{eqn:ms2} into the governing equation Eqn.~\eqref{eqn:gov1} for a plate gives 
\begin{multline}
   \sum_{\substack{p = 1}}^{N_1} \sum_{\substack{q = 1}}^{N_2} \left[{\frac{1}{\sin^4{\theta}}}(\frac{p\pi}{L_1})^4 + {\frac{6}{\sin^2{\theta}\tan^2{\theta}}}(\frac{p\pi}{L_1})^2(\frac{q\pi}{L_2})^2 + \frac{1}{\sin^4{\theta}}(\frac{q\pi}{L_2})^4 + {\frac{2}{\sin^2{\theta}}}(\frac{p\pi}{L_1})^2(\frac{q\pi}{L_2})^2 - \frac{\rho h \omega^2}{D}\right] \cdot \\W_{pq\beta}  \sin{\frac{p \pi x_1}{L_1} \sin{\frac{q \pi x_2}{L_2}}}
   +\frac{4}{\tan{\theta}\sin^3{\theta}}\sum_{\substack{p = 1}}^{N_1} \sum_{\substack{q = 1}}^{N_2}\frac{p \pi}{L_1}\frac{q \pi}{L_2}\left\{(\frac{p \pi}{L1})^2 + (\frac{q \pi}{L2})^2\right\}  W_{pq\beta}  \cos{\frac{p \pi x_1}{L_1} \cos{\frac{q \pi x_2}{L_2}}}
   \\
   = -\frac{k}{D}\sum_{\alpha = 1}^{14 N}\left[\sum_{\substack{p = 1}}^{N_1} \sum_{\substack{q = 1}}^{N_2} W_{pq\beta}  \sin{\frac{p \pi x_1}{L_1} \sin{\frac{q \pi x_2}{L_2}}}-W_{\alpha\beta}\right]
   \delta(\bm{x}-\bm{r}_{\alpha\beta})
   \\
   -\frac{k_{in}}{D}\sum_{\alpha = 1}^{2 N}\left[\sum_{\substack{p = 1}}^{N_1} \sum_{\substack{q = 1}}^{N_2} (W_{pq\beta}-W_{pq\beta'})  \sin{\frac{p \pi x_1}{L_1} \sin{\frac{q \pi x_2}{L_2}}}\right]
   \delta(\bm{x}-\bm{r}_{\alpha\beta}) . 
\end{multline}
Here, $\theta=60^\circ$ is the angle between the two lattice vectors. Multiplying by $\sin(p' \pi x_1/L_1) \sin(q' \pi x_2/L_2)$ and integrating over the lattice gives
\begin{align*}
   &\sum_{\substack{p = 1}}^{N_1} \sum_{\substack{q = 1}}^{N_2} \left[{\frac{1}{\sin^4{\theta}}}(\frac{p\pi}{L_1})^4 + {\frac{6}{\sin^2{\theta}\tan^2{\theta}}}(\frac{p\pi}{L_1})^2(\frac{q\pi}{L_2})^2 +
   \frac{1}{\sin^4{\theta}}(\frac{q\pi}{L_2})^4+{\frac{2}{\sin^2{\theta}}}(\frac{p\pi}{L_1})^2(\frac{q\pi}{L_2})^2 - \frac{\rho h \omega^2}{D}\right] \sin{\theta}W_{pq\beta} \\
   &\int_{0}^{{L_1}} \sin{\frac{p \pi x_1}{L_1} \sin{\frac{p' \pi x_1}{L_1}}}dx_1 \int_{0}^{{L_2}} \sin{\frac{q \pi x_2}{L_2} \sin{\frac{q' \pi x_2}{L_2}}} dx_2
   +\frac{4}{\tan{\theta}\sin^3{\theta}}\sum_{\substack{p = 1}}^{N_1} \sum_{\substack{q = 1}}^{N_2}\frac{p \pi}{L_1}\frac{q \pi}{L_2}\left\{(\frac{p \pi}{L1})^2 + (\frac{q \pi}{L2})^2\right\}  \\
   &\sin{\theta} W_{pq\beta}\int_{0}^{{L_1}} \cos{\frac{p \pi x_1}{L_1} \sin{\frac{p' \pi x_1}{L_1}}}dx_1 \int_{0}^{{L_2}} \cos{\frac{q \pi x_2}{L_2} \sin{\frac{q' \pi x_2}{L_2}}} dx_2\\
   &= -\frac{k}{D}\sum_{\alpha = 1}^{14 N}\sin{\frac{p' \pi x_1}{L_1} \sin{\frac{q' \pi x_2}{L_2}}}\left[\sum_{\substack{p = 1}}^{N_1} \sum_{\substack{q = 1}}^{N_2}W_{pq\beta}  
   \sin{\frac{p \pi x_1}{L_1} \sin{\frac{q \pi x_2}{L_2}}}-W_{\alpha\beta}\right]\iint\delta(\bm{x}-\bm{r}_{\alpha\beta})dx_1 dx_2\\
  & -\frac{k_{in}}{D}\sum_{\alpha = 1}^{2 N}\sin{\frac{p' \pi x_1}{L_1} \sin{\frac{q' \pi x_2}{L_2}}}\left[\sum_{\substack{p = 1}}^{N_1} \sum_{\substack{q = 1}}^{N_2}(W_{pq\beta}-W_{pq\beta'})  
   \sin{\frac{p \pi x_1}{L_1} \sin{\frac{q \pi x_2}{L_2}}}\right]\iint\delta(\bm{x}-\bm{r}_{\alpha\beta})dx_1 dx_2
\end{align*}

Using orthogonality of the basis functions and evaluating the integrals in the above equation, we get 
\begin{align*}
   &\left[{\frac{1}{\sin^4{\theta}}}(\frac{p'\pi}{L_1})^4 + {\frac{6}{\sin^2{\theta}\tan^2{\theta}}}(\frac{p'\pi}{L_1})^2(\frac{q'\pi}{L_2})^2 +
   \frac{1}{\sin^4{\theta}}(\frac{q'\pi}{L_2})^4+{\frac{2}{\sin^2{\theta}}}(\frac{p'\pi}{L_1})^2(\frac{q'\pi}{L_2})^2 - \frac{\rho h \omega^2}{D}\right] \sin{\theta}W_{p'q'\beta}\frac{L_1L_2}{4}\\
   &+\frac{4}{\tan{\theta}\sin^2{\theta}}\sum_{\substack{p = 1}}^{N_1} \sum_{\substack{q = 1}}^{N_2}\frac{p \pi}{L_1}\frac{q \pi}{L_2}\left\{(\frac{p \pi}{L1})^2 + (\frac{q \pi}{L2})^2\right\}  W_{pq\beta}\\ &\frac{L_1L_2}{\pi^2}\left[\frac{\cos{(p-p')\pi}}{q-q'} - \frac{\cos{(q+q')\pi}}{q+q'}\right]\left[\frac{\cos{(q-q')\pi}}{q-q'} - \frac{\cos{(q+q')\pi}}{q+q'}\right]\\
   &=-\sum_{\alpha = 1}^{14 N}\frac{k}{D}\sin{\frac{q' \pi r_{\alpha\beta 1}}{L_1} \sin{\frac{q' \pi r_{\alpha\beta 2}}{L_2}}}\left[\sum_{\substack{p = 1}}^{N_1} \sum_{\substack{q = 1}}^{N_2} W_{pq\beta} 
\sin{\frac{p \pi r_{\alpha\beta 1}}{L_1} \sin{\frac{q \pi r_{\alpha\beta 2}}{L_2}}}-W_{\alpha\beta}\right]\\ &-\sum_{\alpha = 1}^{2 N}\frac{k_{in}}{D}\sin{\frac{p' \pi r_{\alpha 1}}{L_1} \sin{\frac{q' \pi r_{\alpha 2}}{L_2}}}\left[\sum_{\substack{p = 1}}^{N_1} \sum_{\substack{q = 1}}^{N_2} (W_{pq\beta} -W_{pq\beta'})
   \sin{\frac{p \pi r_{\alpha 1}}{L_1} \sin{\frac{q \pi r_{\alpha 2}}{L_2}}}\right] .
\end{align*}
Rearranging the terms in the above equation and substituting $\theta = 60^\circ$ gives Eqn.~\eqref{eqn:ms3}. 


For the resonators, substituting the displacements in Eqns.~\eqref{eqn:ms1} and~\eqref{eqn:ms2} into their governing equations~\eqref{eqn:gov2} gives
\begin{equation}
    -m \omega^2 W_{\alpha\beta}= -k W_{\alpha\beta} + k\sum_{\substack{p = 1}}^{N_1} \sum_{\substack{q = 1}}^{N_2} W_{pq\beta} \sin{\frac{p \pi r_{\alpha\beta1}}{L_1} \sin{\frac{q \pi r_{\alpha\beta2}}{L_2}}} . 
\end{equation}
Dividing both sides by $-m$ gives Eqn.~\eqref{eqn:ms4}

\section{Bloch mode shapes}\label{sec:a_ti}
\begin{figure}[!b]
        \centering     
\includegraphics[height=10cm]{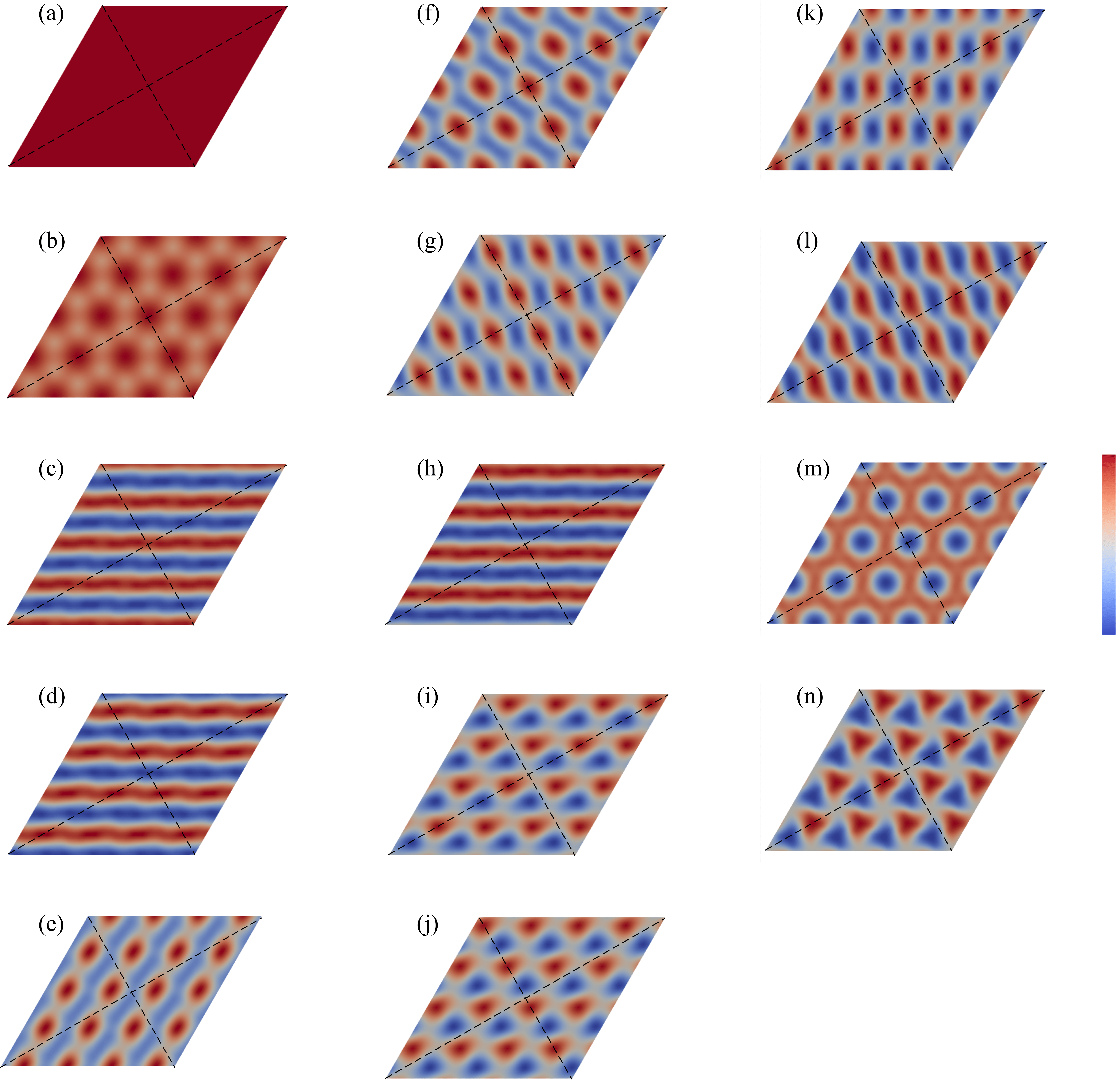}
          \caption{Nonzero projected Bloch mode shapes under rotation by $\theta=\pi$ in the top plate at $\Gamma$ point. Subfigures (a - n) correspond to modes (1 - 14). Modes having rotational eigenvalue  $\lambda_1(p=1)$: 1-7, 13 and $\lambda_2(p=2)$: 8-12, 14. $(4\times4)$ unit cells are shown for clarity of rotational symmetry. }
  \label{fig:ti1}
\end{figure}

\begin{figure}[!hb]
        \centering     \includegraphics[height=10cm]{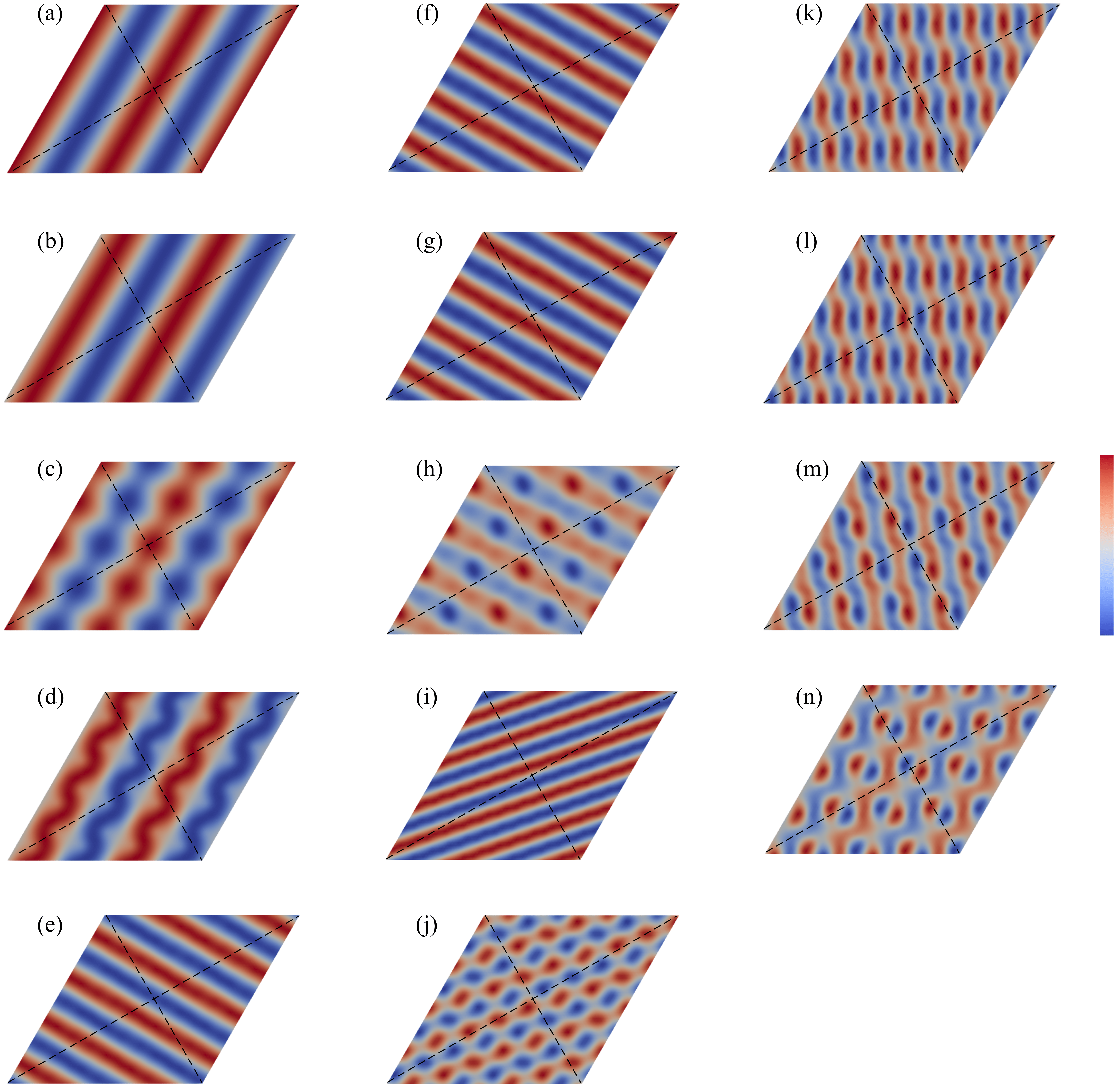}
          \caption{Nonzero projected Bloch mode shapes  under rotation by $\theta=\pi$ in the top plate at $M$ point. Subfigures (a - n) correspond to modes (1 - 14). Modes having rotational eigenvalue $\lambda_1(p=1)$: $1,3,5,8,11,13$ and $\lambda_2(p=2)$: $2,4,6,7,9,10,12,14$. $(4\times4)$ unit cells are shown for clarity of rotational symmetry. 
} 
  \label{fig:ti2}
\end{figure}

\begin{figure}[!t]
        \centering     \includegraphics[height=10cm]{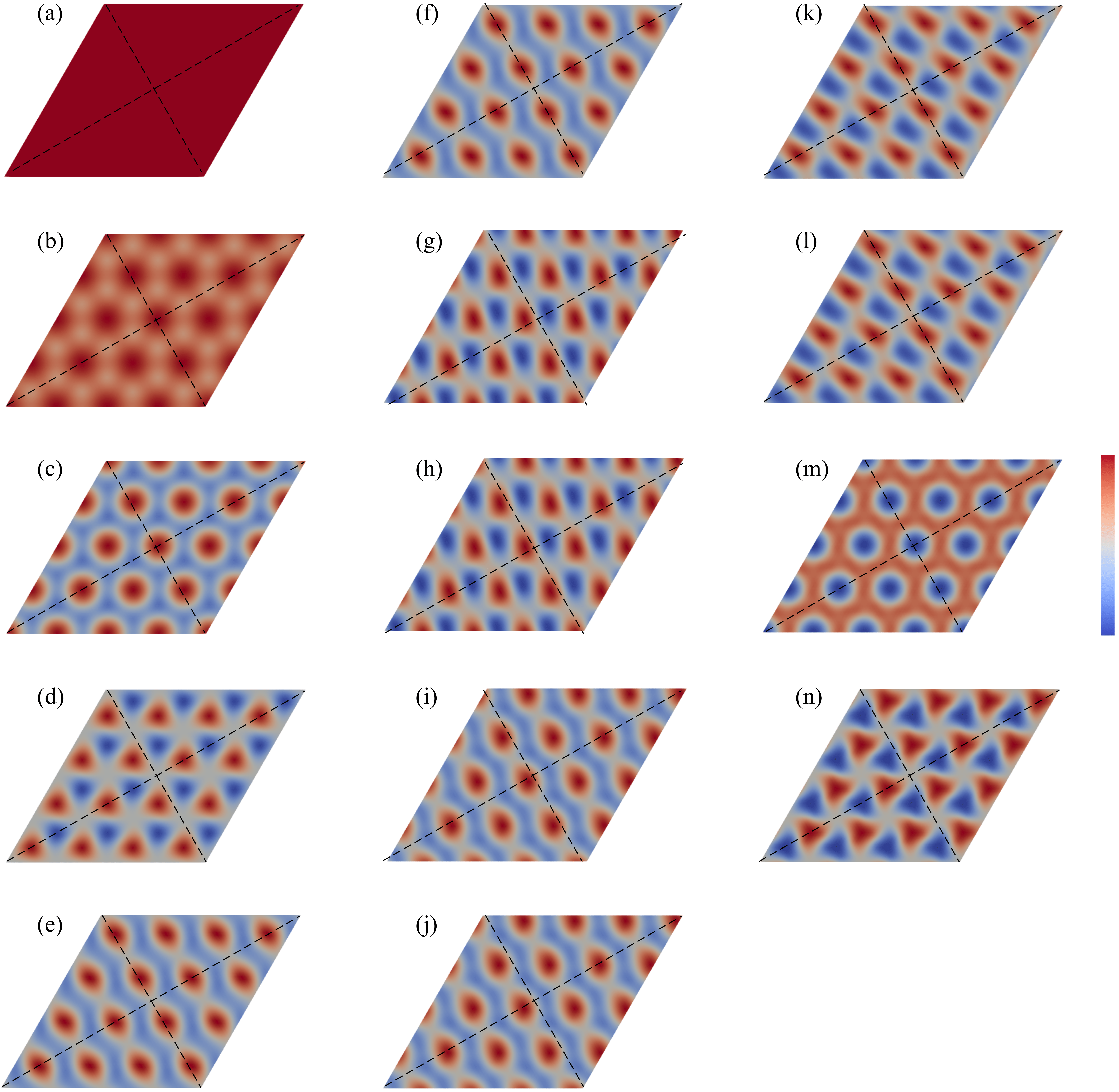}
          \caption{Nonzero projected Bloch mode shapes  under rotation by $\theta=2\pi/3$ in the top plate at $\Gamma$ point. Subfigures (a - n) correspond to modes (1 - 14). Modes having rotational eigenvalue $\lambda_1(p=1)$: $1-4,13,14$, $\lambda_2(p=2)$: $5,7,9,11$ and $\lambda_3(p=3)$: $6,8,10,12$. $(4\times4)$ unit cells are shown for clarity of rotational symmetry. 
          } 
  \label{fig:ti3}
\end{figure}

\begin{figure}[!t]
        \centering     \includegraphics[height=10cm]{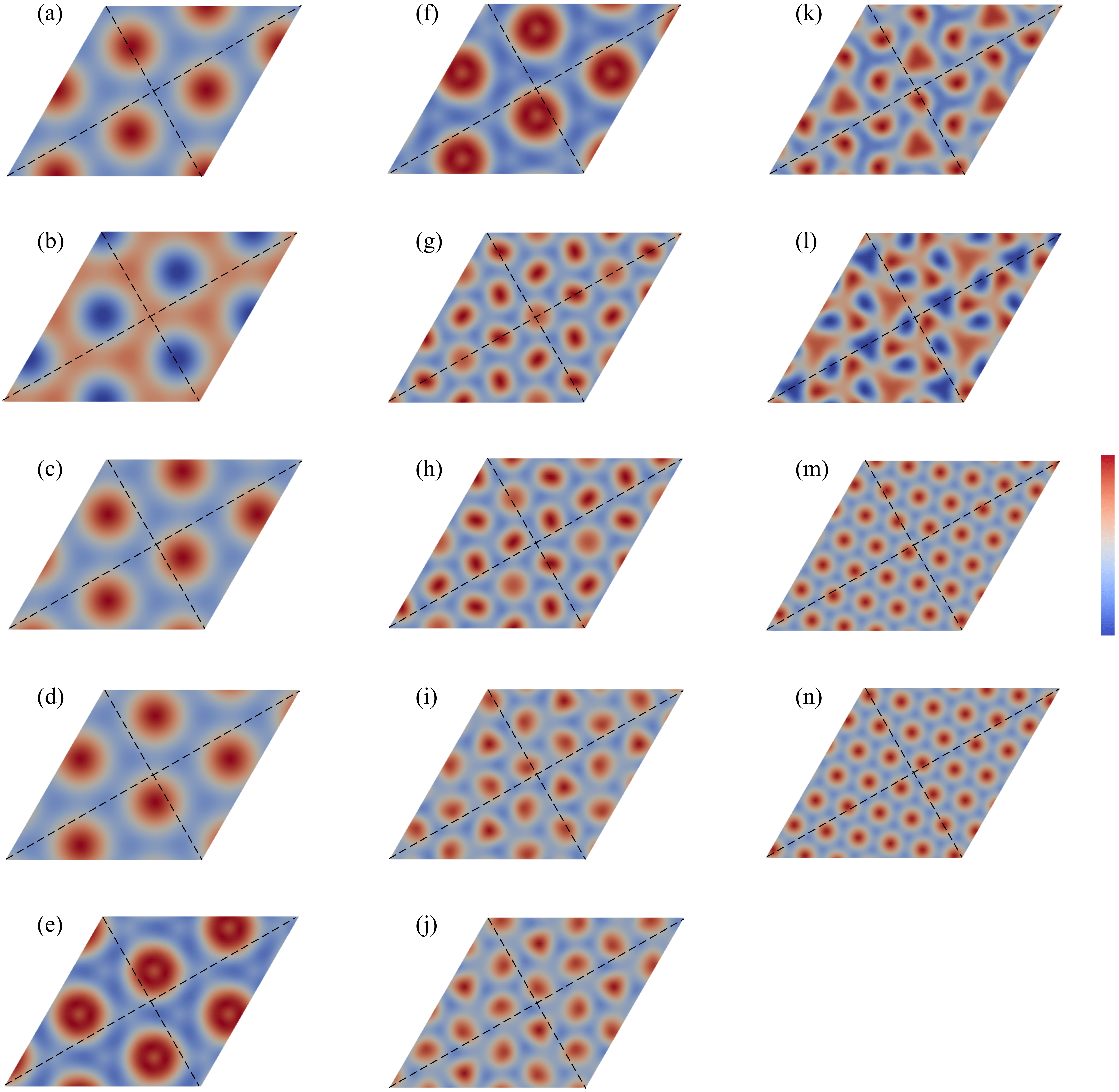}
          \caption{Nonzero projected Bloch mode shapes  under rotation by $\theta=2\pi/3$ in the top plate at $K$ point. Subfigures (a - n) correspond to modes (1 - 14). Modes having rotational eigenvalue $\lambda_1(p=1)$: $1,2,7,8$, $\lambda_2(p=2)$: $3,5,9,11,13$ and $\lambda_3(p=3)$: $4,6,10,12,14$. $(4\times4)$ unit cells are shown for clarity of rotational symmetry. 
} 
  \label{fig:ti4}
\end{figure}

Using the procedure in section~\ref{sec:Qcalc}, the projected Bloch mode shapes $w^p_\beta(\bx)$ for the 14 modes below the band gap are determined at each of the high symmetry points. The non-zero projected mode shapes (real component) are presented in Figures~\ref{fig:ti1} -~\ref{fig:ti4}. Only the top plate is shown although both plates are considered for determining the rotational eigenvalue $\lambda_p$ for each mode. The corresponding $\lambda_p$ for each mode are listed in the captions. Here, multiple unit cells are illustrated for clarity of the rotational symmetry. The rotation axis passes through the center, where the dashed lines intersect. For rotation about this axis by $\theta=\pi$, modes with rotational eigenvalue $\lambda_1$ will be symmetric. Similarly, for rotation by $\theta=2\pi/3$, the mode shapes with $\lambda_1$ will be unchanged after rotation.

\pagebreak

\pagebreak
\clearpage
\section{Primitive generators and their fractional corner modes}\label{sec:a_prim}
\begin{figure}[!htbp]
        \centering     \includegraphics[height=8cm]{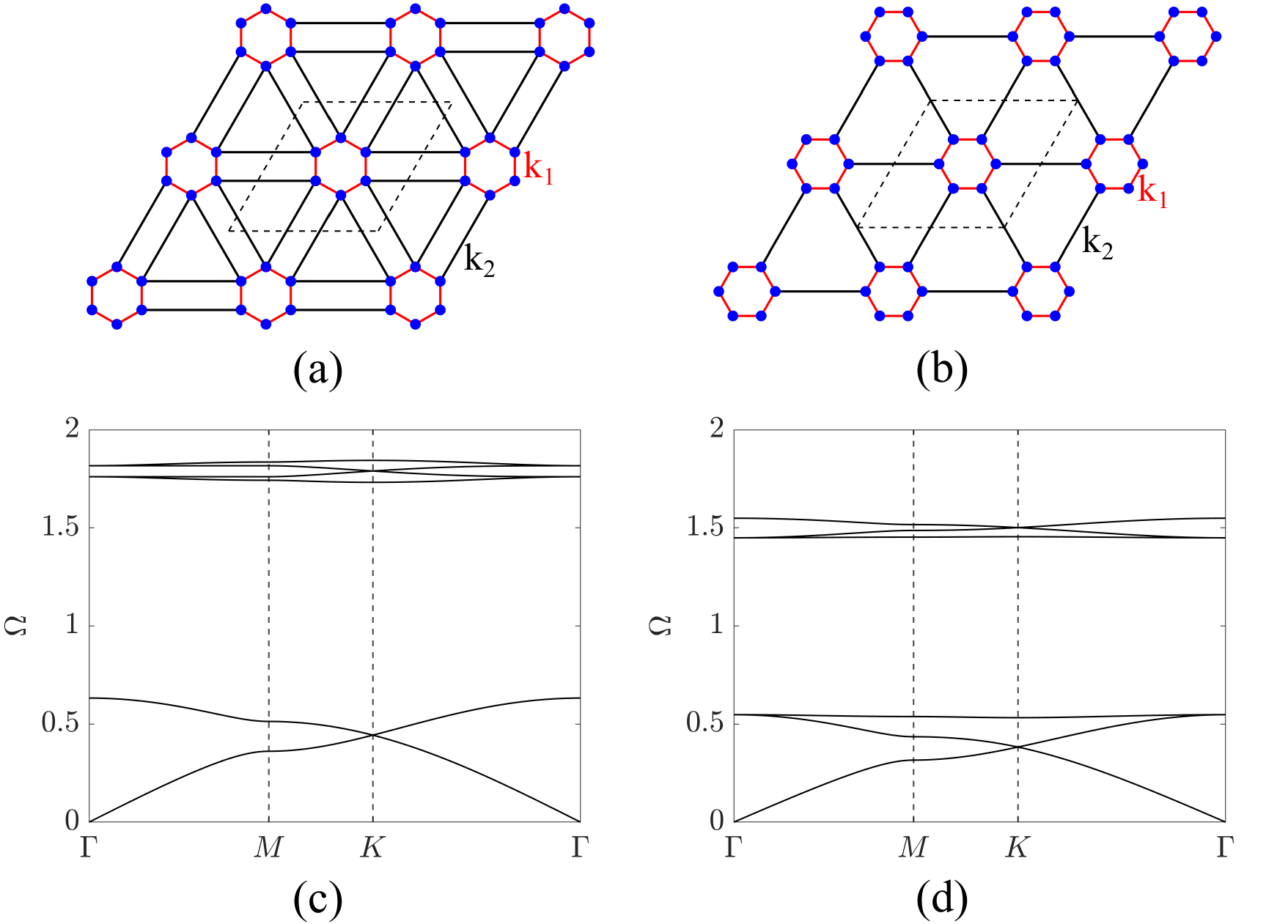}
          \caption{Schematic and dispersion diagram of the primitive generators. Schematic of (a) $h_{4b}$ and (d) $h_{3c}$ lattices. Unit cells are indicated by black dashed lines. Dispersion diagrams are shown along the IBZ boundary for (c) $h_{4b}$ and (d) $h_{3c}$ lattices. 
} 
  \label{fig:a_prim}
\end{figure}

We consider two primitive generators, identical to the ones introduced by Hughes and coworkers~\cite{benalcazar2019quantization}. Figure~\ref{fig:a_prim}(a,c) displays schematics of these lattices. The nodes have point masses with one degree of freedom and can move out-of-plane. The edges have linear springs with stiffness values either $k_1$ or $k_2$ as indicated. In both lattices, a nontrivial topological bandgap opens when $k_1 < k_2$. The dispersion diagrams for these lattices, computed for $k_1=0.1$, $k_2=1.0$ and all unit masses, are displayed in Fig.~\ref{fig:a_prim}(b,d). 

The fractional corner modes at the corners of domains with 6-fold and 3-fold rotation symmetry are  given by~\cite{benalcazar2019quantization}
\begin{subequations}
 \begin{align}
     Q_6&=\dfrac{[M_1^{(2)}]}{4}+\dfrac{[K_1^{(3)}]}{6} \mod 1 
    \label{eqn:a_ti1}
\\
    Q_3&=\dfrac{[K_2^{(3)}]}{3} \mod 1 . 
    \label{eqn:a_ti2}
\end{align}    
\end{subequations}
Here $[K_2^{3}]$ is the difference between the number of modes at $K$ and $\Gamma$ points that have rotational eigenvalue $\lambda_2$. For each mode at the high symmetry points below the bandgap, the rotational eigenvalues are determined using the procedure discussed in Sec.~\ref{sec:Qcalc}. 
For $h_{4b}$ lattice, the fractional corner mode values are 
\begin{subequations}
\begin{align}
    Q_6&=\dfrac{(1-1)}{4}+\dfrac{(0-2)}{6} =-\dfrac{1}{3} \mod 1=\dfrac{2}{3} \mod 1, 
\\
Q_3&=\dfrac{(1-0)}{3} =\dfrac{1}{3} \mod 1,
\end{align}
\end{subequations}
while for $h_{3c}$ lattice, they are
\begin{subequations}
\begin{align}
    Q_6&=\dfrac{(1-3)}{4}+\dfrac{(1-1)}{6} =-\dfrac{1}{2} \mod 1=\dfrac{1}{2} \mod 1
\\
Q_3&=\dfrac{(1-1)}{3} =0.
\end{align}
\end{subequations}

\section{Bulk and localized mode shapes}\label{sec:a_ms}
The complete mode shapes for the modes in Fig.~\ref{fig:ms} are presented. These include the top and bottom plate displacement contours, top and bottom layer resonator displacement contours. 
Note that there are two edge modes at the same frequency. Both mode shapes are illustrated below: subfigures (b, g) for one mode and (c, h) for the other. 
\clearpage
\begin{figure}[!htbp]
        \centering     \includegraphics[height=4.5cm]{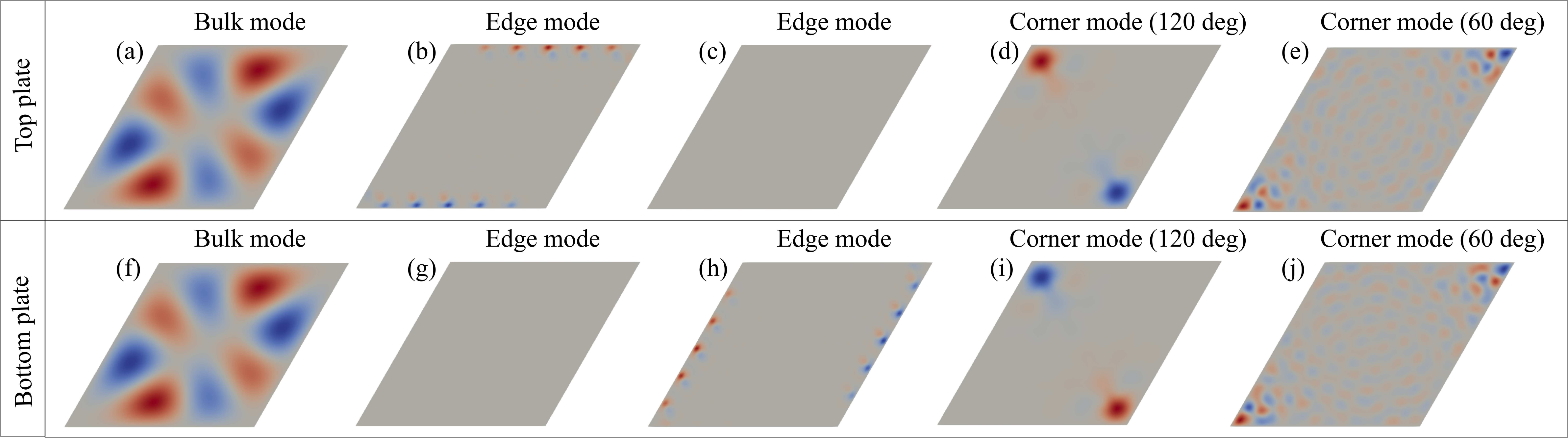}
          \caption{Displacement contours of top and bottom plate for (a, f) a typical bulk mode ($\Omega=0.70$), (b, c, g, h) an edge mode ($\Omega=13.71$), (d, i) corner mode at $120^{\circ}$ corner ($\Omega=2.34$) and (e, j) $60^{\circ}$ corner ($\Omega=8.19$). Top and bottom rows correspond to the top and bottom plate displacement contours, respectively.
} 
  \label{fig:a_ms1}
\end{figure}

\begin{figure}[!htbp]
       \centering     \includegraphics[height=4.5cm]{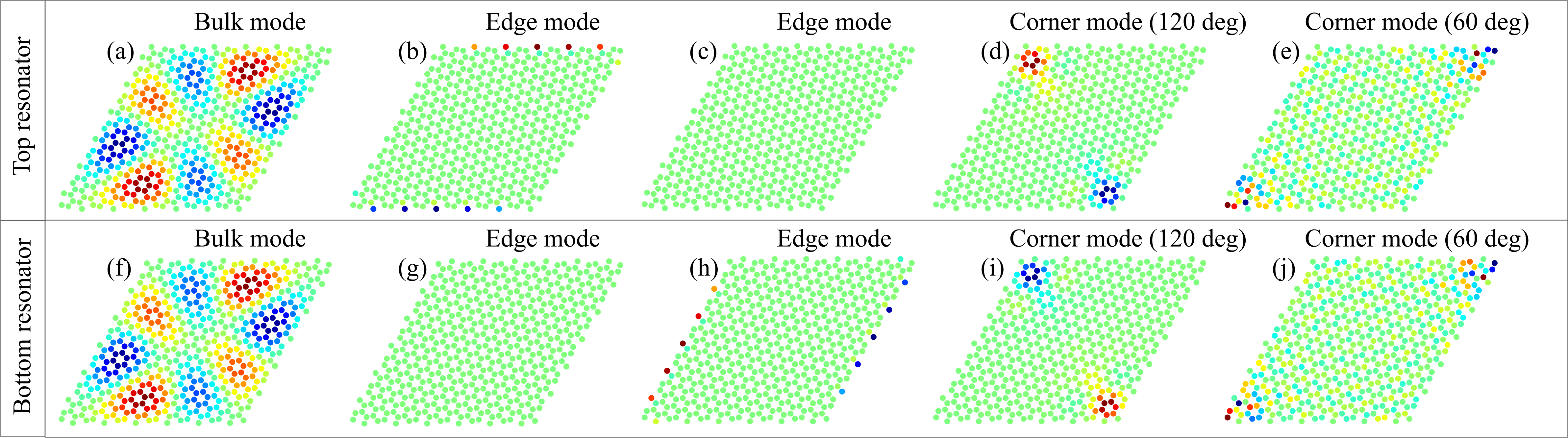}
         \caption{Displacement contours of top and bottom resonator for the same modes presented in Fig.~\ref{fig:a_ms1}. Top and bottom rows correspond to the top and bottom resonator displacement contours, respectively.
} 
 \label{fig:a_ms2}
\end{figure}

\clearpage
\bibliography{apssamp}

\end{document}